%% file: main.tex
\documentclass[runningheads,a4paper]{llncs}

\usepackage[USenglish]{babel}

\usepackage{color}
\usepackage{graphicx}

\usepackage{algorithm}
\usepackage{algorithmic}

\usepackage{amssymb}
\usepackage{amsmath}
\usepackage{stmaryrd}
\usepackage{ltl}

\usepackage{tikz}
\usepackage{pgfplots}
\usepackage{caption}
\usepackage{subcaption}

\usetikzlibrary{positioning}
\usetikzlibrary{calc}
\usetikzlibrary{arrows.meta, automata, positioning, calc}
                
\usepackage{hyperref}
\usepackage[capitalise,nameinlink]{cleveref} 

\usepackage{cite}
\usepackage{hyperref}

\usepackage{paralist}

\input{macros}

\input{examples}
\begin{document}

\mainmatter  

\title{Taming Large Bounds in Synthesis from Bounded-Liveness Specifications (Full Version) \thanks{Philippe Heim carried out this work as PhD candidate at Saarland University, Germany.}}
\titlerunning{Taming Large Bounds in Synthesis from Bounded-Liveness Specifications}

%
%
\author{Philippe Heim \and Rayna Dimitrova }
\authorrunning{}

\institute{CISPA Helmholtz Center for Information Security, Saarbrücken, Germany
\email{{\{philippe.heim, dimitrova\}@cispa.de}}}
%
%

\maketitle

\begin{abstract}
Automatic synthesis from temporal logic specifications is an attractive alternative to manual system design, due to its ability to generate correct-by-construction implementations from high-level specifications. 
Due to the high complexity of the synthesis problem, significant research efforts have been directed at developing practically efficient approaches for restricted specification language fragments.
In this paper we focus on the \safetyLTL\ fragment of Linear Temporal Logic (LTL) syntactically \emph{extended with bounded temporal operators}. 
We propose a new synthesis approach with the primary motivation to solve efficiently the synthesis problem for specifications with bounded temporal operators, in particular those with large bounds. 
The experimental evaluation of our method shows that for this type of specifications it outperforms state-of-art synthesis tools,  demonstrating that it  is a promising approach to efficiently treating quantitative timing constraints in safety  specifications.
\end{abstract}


\section{Introduction}
\input{intro}


\section{Preliminaries}\label{sec:prelim}
\input{preliminaries}

\section{\LTLFrag\ Synthesis with Countdown-Timer Games}\label{sec:timer-games}
\input{timer-games}

\section{Countdown-Timer Game Construction}\label{sec:construction}
\input{tg-generation}

\section{Solving Countdown-Timer Games}\label{sec:solving}
\input{tg-solving}

\section{Evaluation}
\input{evaluation}

\section{Conclusion}
\input{conclusion}

\section*{Data-Availability Statement}

The datasets generated during and/or analysed during the current study are available in the Zenodo repository, \url{https://doi.org/10.5281/zenodo.7505914}.

\bibliographystyle{plain}
\bibliography{bib}

\newpage
\appendix
\input{appendix}

\end{document}

%% file: macros.tex
\newcommand{\power}[1]{2^{#1}}
\newcommand{\sema}[1]{\ensuremath{\llbracket}#1\ensuremath{\rrbracket}}
\newcommand{\NN}{\ensuremath{\mathbb{N}}}

\newcommand{\poly}{\ensuremath{\mathit{poly}}}

\newenvironment{arrayeq}[1]{\begin{equation*}\begin{array}{#1}}{\end{array}\end{equation*}}

\newcommand{\lnext}{\LTLnext} 
\newcommand{\blnext}[1]{\LTLnext[#1]}
\newcommand{\eventually}{\LTLfinally} 
\newcommand{\beventually}[1]{\LTLfinally[#1]}
\newcommand{\globally}{\LTLglobally} 
\newcommand{\bglobally}[1]{\LTLglobally[#1]}
\newcommand{\until}{\LTLuntil}
\newcommand{\buntil}[1]{\LTLuntil[#1]}
\newcommand{\weak}{\LTLweakuntil}
\newcommand{\bweak}[1]{\LTLweakuntil[#1]}

\newcommand{\timers}{\ensuremath{\mathcal{T}}}
\newcommand{\effect}{\ensuremath{\mathcal{E}}}
\newcommand{\timerspace}{\ensuremath{\mathcal{V}}}
\newcommand{\reset}{\ensuremath{\mathit{RESET}}}
\newcommand{\unsafe}{\ensuremath{\mathit{UNSAFE}}}

\newcommand{\LTLFrag}{\ensuremath{\mathit{SafeLTL}_B}}
\newcommand{\LTLFragT}{\ensuremath{\mathit{SafeLTL}_B^{t}}}
\newcommand{\LTL}{LTL}
\newcommand{\safetyLTL}{\texttt{Safety LTL}}
\newcommand{\LTLEBR}{\ensuremath{\text{LTL}_\mathsf{EBR}}}

\newcommand{\PLTL}{PLTL}
\newcommand{\promptLTL}{{\sc prompt}-LTL}

\newcommand{\rep}{\mathit{Rep}}

\newcommand{\states}{\ensuremath{S}}

\newcommand{\outcome}[3]{\mathit{Outcome}(#1,#2,#3)}

\newcommand{\exptime}{\texttt{EXPTIME}}

\newcommand{\myparabf}[1]{\smallskip\noindent{\bf #1}}
\newcommand{\myparait}[1]{\smallskip\noindent{\it #1}}

\newcommand{\office}{\mathit{office}}
\newcommand{\corridor}{\mathit{corridor}}
\newcommand{\charge}{\mathit{charge}}
\newcommand{\human}{\mathit{human}}
\newcommand{\night}{\mathit{night}}
\newcommand{\request}{\mathit{request}}
\newcommand{\makecoffee}{\mathit{makeCoffee}}
\newcommand{\release}{\mathit{release}}
\newcommand{\stuck}{\mathit{stuck}}
\newcommand{\resume}{\mathit{resume}}
\newcommand{\move}{\mathit{move}}
\newcommand{\stopp}{\mathit{stop}}
\newcommand{\con}{\mathit{on}}
\newcommand{\coff}{\mathit{off}}
\newcommand{\pick}{\mathit{pick}}
\newcommand{\pput}{\mathit{put}}
\newcommand{\pin}{\mathit{in}}
\newcommand{\travel}{\mathit{travel}}
\newcommand{\closed}{\mathit{closed}}
\newcommand{\opened}{\mathit{opened}}
\newcommand{\transit}{\mathit{transit}}
\newcommand{\popen}{\mathit{open}}
\newcommand{\pclose}{\mathit{close}}

\newcommand{\req}{\mathit{r}}
\newcommand{\gr}{\mathit{g}}

%% file: intro.tex
Reactive synthesis~\cite{Church62} has the goal of automatically generating an implementation from a formal specification that describes the desired behavior of a reactive  system. 
The system requirements are typically specified using temporal logics such as Linear Temporal Logic (\LTL). Temporal logics are expressive, high-level specification languages capable of describing rich properties, such as, for example,  robotic missions~\cite{Kress-GazitFP09}.
Specifications of  reactive systems often include requirements of the form 
``something good eventually happens''.  These can be expressed in \LTL\ via the temporal operators $\until$~(``until") and $\eventually $~(``eventually"). 
``Eventually'' is an abstraction for the existence of some unknown time point in the future of a system execution when some property holds true. 
While this abstraction is useful for avoiding over-specification,  there are many situations in which there are practical bounds on the time within which a  requirement must be met.  
In such cases, it is vital that the synthesis procedure checks if the timing requirements are realizable, and synthesizes an implementation that adheres to these  bounds.  
\looseness=-1

As a simple example,  consider a specification of the desired behavior of a controller for the front door of an office building.  
Our specification states that the door must always be locked at night, and unlocked otherwise. 
It also stipulates that in the event of a fire the door should eventually open.
Formulated like this, the specification is realizable.  
However,  in case of a fire during night the synthesized implementation will only open the door at the start of the day. Clearly,  this is not the behavior we intended! 
We can specify the actual desired behavior in \LTL\ by using the temporal operator $\lnext$~(``next"), which allows us to state that a property should hold at the next time step. However,  we would need to use nested $\lnext$ operators in order to express the required time bounds. This can quickly become inconvenient,  especially if we need to specify various different time bounds,  some of them large. 
This modeling inconvenience and the increase of specification size are easily avoided by adding bounded versions of the temporal operators as syntactic sugar, without increasing  expressiveness.

Due to their practical significance,  fragments of \LTL\ in which the formulas (in negation normal form) include only bounded versions of the $\until$ and $\eventually $ operators have attracted considerable attention. 
The most prominent such fragment is \safetyLTL\, the until-free fragment of \LTL\ in negated normal form.
Since \safetyLTL\  is a syntactic fragment of \LTL, it can express bounded liveness properties only via nested next operators.
Another notable example is the logic Extended Bounded Response LTL (\LTLEBR)~\cite{CimattiGGMT20}, which is a fragment of \LTL\ that includes bounded temporal operators as well as unbounded universal temporal operators (i.e., ``globally'' and ``release'').
While every \LTLEBR\ formula can be expressed in \safetyLTL, one significant advantage of \LTLEBR\ is that the bounds of the temporal operators are represented in binary,  which allows for exponentially more succinct formulas.
However,  in the course of the synthesis procedure presented in~\cite{CimattiGGMT20} these bounds are expanded into nested ``next'' operators. Keeping  bounds symbolic is identified in~\cite{CimattiGGMT20} as an interesting direction for future developments. Indeed, in many practically relevant cases large bounds are unavoidable due to requirements on the same system across different time-scales. 

In this paper we address this challenge by proposing a synthesis procedure for an extension of \safetyLTL\  with bounded operators. We develop dedicated techniques for handling  the temporal bounds symbolically and efficiently. 



\myparabf{Contribution.} 
We propose a synthesis method for specifications expressed in a fragment of \LTL\ which is a syntactic extension of \safetyLTL\ with bounded temporal operators.  
The distinguishing characteristic of our method is a reduction to a dedicated game model, called \emph{countdown-timer games} in which the temporal operators' bounds are treated symbolically via the introduction of \emph{timers}.  
Further features of the translation are techniques for on-the-fly pruning of edges in the constructed game and reduction of the number of introduced timers.  
We present an abstraction-based method for solving the resulting games.
We have developed a prototype implementation of our approach,  and the experimental evaluation demonstrates that it is indeed capable of handling efficiently safety specifications with large bounds.
We demonstrate that on a set of benchmarks featuring bounded temporal operators with large bounds, our technique outperforms state-of-the-art tools for \LTLEBR\  and \LTL\ synthesis.  

\myparabf{Related Work.} 
\input{related-work}

%% file: related-work.tex
\emph{The synthesis problem for  \safetyLTL} has attracted significant interest due to its algorithmic simplicity compared to general \LTL\ synthesis~\cite{ZhuTLPV17}.  
For instance,  the symbolic approach presented in~\cite{ZhuTLPV17} is shown to outperform the state-of-the-art \LTL\ synthesis tools at the time.
For \LTLEBR,~\cite{CimattiGGMT20} proposes a synthesis algorithm based on a fully symbolic translation to deterministic  safety automata.  
A key difference between our approach and the above techniques is that our countdown-timer game construction does not expand upfront the bounded temporal operators, but treats them symbolically instead. 
Furthermore,  the authors of~\cite{ZhuTLPV17}  point out that for large \safetyLTL\ formulas the construction of the deterministic safety automaton presents a performance bottleneck. 
Our safety game constriction makes use of pruning in order to alleviate this problem by eliminating on-the-fly parts of the game graph that need not be explored.

\emph{Parameterized temporal logics}, such as \PLTL~\cite{AlurETP01}  enable the specification of parametric lower and upper bounds on the satisfaction time of the ``globally" operator and the wait time of ``eventually".  
In the logic \promptLTL~\cite{KupfermanPV09}, only eventualities are parameterized by upper bounds. 
%
The bounds of the temporal operators in these logics are unknown parameters,  while in the case that we consider, the bounds are given integer constants.  The goal of our work is to develop a synthesis method that treats constant bounds efficiently. 

In the \emph{real-time setting}, temporal logics that allow for limiting the time scope of temporal operators have been extensively studied.  Notable logics are Metric Temporal Logic (MTL)~\cite{Koymans90},  and its fragment Metric Interval Temporal Logic (MITL)~\cite{AlurFH96}.
Compared to the untimed setting,  synthesis from real-time logic specifications poses additional challenges.
Controller synthesis is undecidable for MTL~\cite{BouyerBC06},  for MITL~\cite{DoyenGRR09 , BrihayeEGHMS16},  and even for the safety fragment of MTL~\cite{BrihayeEGHMS16}. 
Decidability is regained by fixing the resources (clocks and  guards) of the controller~\cite{DSouzaM02 , BrihayeEGHMS16}.  
The key challenge stems from the fact that synthesis requires deterministic automata, and it is not generally possible to construct deterministic timed automata for MITL.
To circumvent this problem,  the assumption of bounded variability
  is commonly made.
Under this assumption,~\cite{MalerNP07} proposes a synthesis algorithm for bounded response properties, and a translation from
MTL to deterministic timed automata 
is presented in~\cite{NickovicP10}. 
With respect to tool support,  sound but incomplete synthesis methods for fragments of MTL have been proposed in~\cite{BulychevDLL14} and~\cite{LiJLLP17}, and implemented in toolchains that employ  {\sc Uppaal-Tiga}~\cite{BehrmannCDFLL07} for timed games solving.
A tool for MTL controller synthesis via translation to alternating timed automata was presented in~\cite{HofmannS21}.
In the case when the real-time synthesis problem is given as a timed game
and the specification is  a state-based winning condition, 
the problem of computing a control strategy is decidable~\cite{MalerPS95}.  Efficient on-the-fly algorithms for timed games have been developed~\cite{Cassez07}, and successfully implemented in {\sc Uppaal-Tiga}~\cite{BehrmannCDFLL07} and  {\sc Uppaal-Stratego}\cite{DavidJLMT15}.\looseness=-1
 Since we are interested in discrete-time systems, we circumvent the additional challenges present in the dense-time setting by remaining the realm of discrete time and focusing on efficiently treating quantitative timing constraints there.

%% file: preliminaries.tex
\myparabf{Reactive Synthesis}
Let $\mathcal{I}$ be a finite set of uncontrollable environment \emph{input Boolean propositions} and $\mathcal{O}$ be a finite set of controllable \emph{output Boolean propositions}.
A \emph{reactive system} is a tuple $(C, c_0, \gamma)$ where $C$ is a set of \emph{control states}, 
$c_0 \in C$ the \emph{initial control state},  and 
$\gamma: C \times \power{\mathcal{I}} \to C \times \power{\mathcal{O}}$  is the \emph{transition function}.
A \emph{specification} is a language $\mathcal{L} \subseteq {\left(\power{\mathcal{I} \cup \mathcal{O}} \right)}^\omega$ of infinite words over $\mathcal I \cup \mathcal O$.

A system $(C, c_0, \gamma)$ \emph{realizes} a specification $\mathcal L$ if for all infinite sequences of environment inputs $i \in {\left(\power{\mathcal{I}}\right)}^\omega$ it yields an output sequence $o \in {\left(\power{\mathcal{O}}\right)}^\omega$ defined by $(c_{t+1}, o_{t}) = \gamma(c_t, i_t)$ for $t \in \NN$, such that $i \cup o \in \mathcal{L}$. 
\emph{Reactive synthesis} is the problem of finding a realizing implementation for a given specification.

\myparabf{Safety LTL with Bounded Liveness Operators}
We consider specifications expressed using temporal logic, more concretely, in a fragment of LTL~\cite{Pnueli77}, which we denote by \LTLFrag.
The fragment \LTLFrag\ is a syntactic extension of \safetyLTL~\cite{ZhuTLPV17} and defined by the following grammar:
    \[ \varphi, \psi := \mathit{ap} \;|\; \lnot \mathit{ap} \;|\; \varphi \land \psi \;|\; \varphi \lor \psi  \;|\; \blnext{n} \varphi \;|\; \beventually{n} \varphi \;|\;  \varphi \bweak{n} \psi \;|\; \varphi \weak \psi \]
for $\mathit{ap} \in \mathcal{I} \cup \mathcal{O}$ and $n \in \NN$. \looseness=-1
\LTLFrag\ extends \safetyLTL\ by bounded operators with  bounds encoded in binary.
While all bounded operators have equivalent \safetyLTL\ formulas (e.g. $\beventually{n} \varphi \equiv \bigvee_{i \in \{0 \dots n\}} \lnext^i \varphi$) these have exponentially larger encoding.
The constants $\top$~(true),  $\bot$~(false),  the ``globally'' operator $\globally$ and ``bounded until'' $\buntil{n}$ can be derived as 
$\top := a \lor \lnot a$, 
$\bot := a \land \lnot a$, 
$\globally \varphi := \varphi \weak \bot$, 
$\bglobally{n} \varphi := \varphi \bweak{n} \bot$,  and
$\varphi\buntil{n} \psi := (\varphi \bweak{n} \psi) \wedge \beventually{n} \psi$,
respectively.

The satisfaction of  a formula $\Phi \in \LTLFrag$ by infinite word $w=w_0w_1\ldots \in {\left(\power{\mathcal{I} \cup \mathcal{O}}\right)}^\omega$ at time point $k \in \NN$ is denoted as $w \vDash_k \Phi$ and is defined follows:
\begin{arrayeq}{lllll}
          w \vDash_k a &:\Leftrightarrow a \in w_k
& \quad & w \vDash_k \lnot a &:\Leftrightarrow a \not\in w_k \\
          w \vDash_k \varphi \land \psi &:\Leftrightarrow (w \vDash_k \varphi) \land (w \vDash_k \psi)
& \quad & w \vDash_k \varphi \lor \psi &:\Leftrightarrow (w \vDash_k \varphi) \lor (w \vDash_k \psi) \\
          w \vDash_k \beventually{n} \varphi &:\Leftrightarrow \exists i \leq n.~ w \vDash_{k + i} \varphi 
& \quad & w \vDash_k \blnext{n} \varphi &:\Leftrightarrow w \vDash_{k + n} \varphi 
\end{arrayeq}
\vspace{-0.29cm}
\begin{arrayeq}{ll}
    w \vDash_k \varphi \bweak{n} \psi &:\Leftrightarrow (\forall i \leq n. w \vDash_{k+i} \varphi) \lor (\exists j \leq n. w \vDash_{k+j} \psi \land \forall i < j. w \vDash_{k+i} \varphi) \\
    w \vDash_k \varphi \weak \psi &:\Leftrightarrow (\forall i. w \vDash_{k+i} \varphi) \lor (\exists j. w \vDash_{k+j} \psi \land \forall i < j. w \vDash_{k+i} \varphi).
\end{arrayeq}
The language of $\Phi \in \LTLFrag$ is defined as $\mathcal{L}(\Phi) := \{ w \in {\left(\power{\mathcal{I} \cup \mathcal{O}}\right)}^\omega \mid w \vDash_0 \Phi \}$.

\myparabf{Two-Player Safety Games}
The synthesis problem for temporal logic specifications can be solved by translating the specification into a two-player game between the system and the environment, and then solving the game to determine the winning player.  If the system wins,  an implementation can be extracted.

A \emph{game structure} is a tuple $G = (\states, \states_0, \mathcal{I}, \mathcal{O}, \rho)$, where 
$\states$ is a \emph{set of states}, 
$\states_0 \subseteq \states$ is a set of \emph{initial states},
$\mathcal{I}$ and $\mathcal{O}$ are sets of propositions as defined earlier, and
$\rho: \states \times \power{\mathcal{I}} \times  \power{\mathcal{O}} \to \states$ is a \emph{transition function}. 
A game on $G$ is played by two players,  the system and the environment.
In a given state $s\in \states$, 
the environment chooses some input $i \subseteq \mathcal{I}$, then
the system chooses some output $o \subseteq \mathcal{O}$, and
these choices determine the next state $s' := \rho(s, i, o)$.
The game then continues from $s'$.
The resulting infinite sequence $\pi = s_0,s_1,s_2,\ldots$ of states is called a \emph{play}.
Formally, a play is a sequence $\pi = s_0,s_1,s_2,\ldots \in \states^\omega$ such that $s_0 \in S_0$ and for every $t \in \mathbb N$, $s_{t+1} = \rho(s_t,i,o)$.
A \emph{system strategy} is a function $\sigma: \states^+ \times \power{\mathcal{I}} \to \power{\mathcal{O}}$. An \emph{environment strategy} is a function $\pi : \states^+ \to  \power{\mathcal{I}}$.
Given a state $s \in \states$, a system strategy $\sigma$ and an environment strategy $\pi$, we denote with $\outcome{s}{\pi}{\sigma}$ the unique play $s_0,s_1,s_2,\ldots$ such that $s_0 = s$, and for all $k \in \mathbb N$, $s_{k+1} = \rho(s_k, i_k, \sigma((s_0,s_1\ldots,s_k), i_k))$, where 
$i_k  = \pi((s_0,s_1\ldots,s_k))$.

A \emph{safety game} is a tuple $(G, \mathit{UNSAFE})$ where $\mathit{UNSAFE} \subseteq S$ are unsafe states. 
The system wins the safety game if it has a strategy $\sigma$ such that for all environment strategies $\pi$, $s_0 \in S_0, k \in \NN$, it holds that $\outcome{s_0}{\pi}{\sigma}_k \not\in \mathit{UNSAFE}$.
Such strategy is called a \emph{winning strategy} for the system.
Intuitively, the system has to avoid the unsafe states no matter what the environment does.
The environment wins if it can enforce a visit  to $\unsafe$, i.e.,  when there exist environment strategy $\pi$ and $s_0 \in S_0$ such that for every system strategy $\sigma$ there exists $k \in \NN$ such that $\outcome{s_0}{\pi}{\sigma}_k \in \mathit{UNSAFE}$.


%% file: timer-games.tex
\myparabf{$\boldsymbol{\LTLFrag}$ Synthesis}
We consider the realizability and synthesis problems for the fragment $\LTLFrag$. 
We  focus on the challenge of handling efficiently specifications with large bounds in the bounded temporal operators, and propose a new synthesis method towards achieving this goal.
The proposed approach proceeds in two stages.
In the first stage, the given \LTLFrag\ formula is transformed into a kind of safety game, in which bounds are treated symbolically.  We term these games \emph{countdown-timer games},  introduced later in this section.
The second stage of our synthesis algorithm is the solving of the generated countdown-timer game in order to determine the winning player and answer the realizability question.
We propose in Section~\ref{sec:solving} a method that employs symbolic representation and approximations in order to efficiently solve such games in practice.
\looseness=-1

\myparabf{Countdown-Timer Games}
Intuitively, countdown-timer games are like safety games but with additional \emph{countdown-timers}. 
Countdown-timers are discrete timers that always start with an assigned duration and are decremented by one with every transition in the game.
Once a timer reaches zero it times out, and the transition relation of the countdown-timer game may depend on this information for determining the successor state.  
A countdown-timer can be reset to the duration associated with it.
In addition, countdown-timers with the same duration can swap their values, which we will later use when generating timer-games to avoid unnecessary blowup in the number of timers.

\begin{definition}[Countdown-Timer Games]\label{def:timergames}
A countdown-timer game structure is a tuple $G_T = (\timers, d, L, L_0, \mathcal{I}, \mathcal{O}, \delta)$ where 
\timers\ is a finite set of \emph{countdown timers}, 
$d: \timers \to \NN$ associates a \emph{duration} with each timer, 
$L$ is a finite set of game locations, 
$L_0 \subseteq L$ is the set of initial locations, 
$\mathcal{I}$, $\mathcal{O}$ are finite sets of  uncontrollable environment input propositions and controllable system propositions, respectively, and  
$\delta: L \times \power{\mathcal{I}} \times \power{\mathcal{O}} \times \power{\timers} \to L \times \effect$ is the transition relation. 
$\effect := \timers \to \left( \timers \cup \{\reset\} \right)$ is the \emph{set  of effects} where for all $e \in \effect$:
\begin{compactenum}
    \item\label{cond:effect-1}   for all $t \in \timers$ either $e(t) = \reset$, or $e(t) \in \timers$ and $ d(e(t)) = d(t)$ and,
    \item\label{cond:effect-2}   for $t_1, t_2 \in \timers$ with $t_1 \neq t_2$ we have $e(t_1) \neq e(t_2)$ or $e(t_1)= e(t_2)=\reset$.
\end{compactenum}
A countdown-timer game is a pair $(G_T, \mathit{UNSAFE}_L)$ where $\mathit{UNSAFE}_L \subseteq L$ is a set of \emph{unsafe locations}.
\end{definition}
The effects $\effect$ capture the resets and remapping of timers that can occur upon transitions.
Condition~(\ref{cond:effect-1}) states that each timer is either reset or remapped to a timer with the same duration.
Condition~(\ref{cond:effect-2}) requires the remapping to be injective, i.e.\ no two timers are mapped to the same timer.
When timers are not reset and not remapped to other timers, they are simply mapped to themselves.

The semantics of a countdown-timer game is the safety game generated by explicitly expanding the possible valuations of the timers. 
Intuitively, each state of the game structure is a pair $s = (l,v)$ of a location $l \in L$ and a timer valuation $v$. 
Initially, each timer $t$ is set to its associated duration $d(t)$. 
The transition relation updates the values of the timers by first decrementing them and then applying the effect $e$ of the corresponding transition in $G_{T}$.  
The relevant transition in $G_{T}$ is determined by the location $l$, the input and output sets $i$ and $o$,  and the set of timers whose value has become $0$ after the decrementation.
\begin{definition}[Countdown-Timer Games Semantics]\label{def:timer-game-semantics}
In the context of Definition~\ref{def:timergames}, let 
$\timerspace := \{ v : \timers \to \NN \mid \forall t \in \timers.~v(t) \leq d(t) \}$ 
be the space of all possible timer valuations.
Let $G = (L \times \timerspace, L_0 \times \{ \lambda t.d(t) \}, \mathcal{I}, \mathcal{O}, \rho)$ be a game structure where $\rho ((l,v), i, o) := \mathit{trans}(l, \mathit{step}(v), i, o)$ with 
\begin{arrayeq}{rl}
    \mathit{step}(v) & := \lambda t. \max \{0, v(t) - 1\} \\
    \mathit{trans}(l, v, i, o) &:= \begin{cases}
    \left(l', 
        \lambda t.~\begin{cases}
                v(e(t)) &~\text{if}~e(t) \in \timers \\
                d(t)    &~\text{if}~e(t) = \reset
    \end{cases}\right),\\
            \text{where}~(l', e) := \delta(l, i, o, \{ t \in \timers \mid v(t) = 0\}). \\
    \end{cases}
\end{arrayeq}
The semantics of the countdown-timer game $(G_T, \mathit{UNSAFE}_L)$ is the safety game 
$(G, \mathit{UNSAFE_L} \times \timerspace)$. 
The system (environment) wins the countdown-timer game if and only if it wins the safety game representing its semantics.
\end{definition}

%% file: tg-generation.tex
We now present the first phase of our synthesis algorithm, namely the translation of a \LTLFrag\ formula  to a countdown-timer game.
Our construction is based on expansion rules.
For example, the formula $\beventually{50} a$ is equivalent to $a \lor \lnext\beventually{49} a$.
If $a$ is true, then the whole formula is true.
Otherwise, in the next step $\beventually{49} a$ has to hold.
Interpreted as a state of a safety game, $\beventually{50} a$ has a transition to $\top$ on $a = \top$ and to $\beventually{49} a$ on $a = \bot$. 
This can be repeated on $\beventually{49} a$ and so on.
Once we reach $\beventually{0} a$ we expand it to $a \lor \lnext\bot$, and hence, $a=\bot$ leads to $\bot$ which is the unsafe state.
This construction works for safety formulas, as rejection can be decided with a finite prefix.
As we show later, generating a game structure in this way has the advantage that it can be pruned using information from the formula.
\looseness=-1

However,  this explicit expansion yields a sequence of formulas that is linear in the bound, and hence, exponential in the description of the formula.
Instead of explicit bounds,  we use countdown-timers representing multiple values. 
In the above example, we do not generate all the expansions $\beventually{50} a$, \dots,  $\beventually{0} a$,  but instead a timer $t$ with duration 51 to represent all expansions from 50 to 0 in the single location $a \lor \beventually{t} a$.
If $t$ times out, $\beventually{t}$ has reached the end of the expansion and is transformed to $\bot$.
Hence, instead of having $\beventually{50} a$, \dots, $\beventually{0} a$, $\top$ and $\bot$ as states of a safety game we only have locations $a \lor \beventually{t} a$, $\top$ and $\bot$ in a countdown-timer game. 
We now describe this construction formally.

\subsection{Construction of a Countdown-Timer Game from $\LTLFrag$}

The locations of the generated countdown-timer games are $\LTLFrag$ formulas with, additionally, timers as bounds of the temporal operators.
We denote the set of these formulas as $\LTLFragT$.
Given a set of timers $\timers$, the grammar of $\LTLFragT$ is the grammar of $\LTLFrag$ but in $\beventually{n}$, $\blnext{n}$, and $\bweak{n}$ we have $n \in \NN \cup \timers$.
For $\varphi \in \LTLFragT$, $\mathit{Timers}(\varphi) \subseteq \timers$ denotes all timers appearing in $\varphi$.
\looseness=-1

\myparabf{Game Structure}
Let $\Phi$ be a $\LTLFrag$ formula over input propositions $\mathcal{I}$ and output propositions $\mathcal{O}$.
We construct a countdown-timer game structure $(\timers, d, L, L_0, \mathcal{I}, \mathcal{O}, \delta)$ as follows.  The set of timers
\[\timers := \{ t_i^d \mid \blnext{d}, \beventually{d-1}, \text{ or } \bweak{d-1} \text{ occurs in } \Phi, 0 \leq i \leq d \}\] 
consists of timers $t_i^d$ with index $i$ and durations
$d(t_i^d) := d$ for $0\leq i \leq d$.
The duration of a timer determines the bounds of the temporal operators in $\Phi$ for which it can be used,  and  the indices are used for distinguishing multiple timers of the same duration (introduced at different points of the expansion).  

Let $L := \mathit{PositiveBooleanCombinations}(\mathit{cl}(\Phi))$ (i.e.,  built from $\mathit{cl}(\Phi)$ using $\land,\lor$) be the set of locations, where $\mathit{cl}$ is the \emph{closure} operator
defined as:
\begin{arrayeq}{lll}
    \mathit{cl}(l) & := \{l, \top, \bot \} & l \in \{ \mathit{ap}, \lnot\mathit{ap}\}\\ 
    \mathit{cl}(\varphi~o~\psi) & := \mathit{cl}(\varphi) \cup \mathit{cl}(\psi) & o \in \{\land, \lor\}\\ 
    \mathit{cl}(\blnext{n}\varphi)  & := \mathit{cl}(\varphi) \cup \{ \blnext{t_i^n}\varphi \mid  0 \leq i \leq n \} \\
    \mathit{cl}(\beventually{n}\varphi)  & := \mathit{cl}(\varphi) \cup \{ \beventually{t_i^{n + 1}}\varphi \mid  0 \leq i \leq n + 1 \} \\
    \mathit{cl}(\varphi \bweak{n} \psi)  & := \mathit{cl}(\varphi) \cup \mathit{cl}(\psi) \cup \{ \varphi \bweak{t_i^{n + 1}} \psi \mid  0 \leq i \leq n + 1\} \\
    \mathit{cl}(\varphi \weak \psi)  & := \mathit{cl}(\varphi) \cup \mathit{cl}(\psi) \cup \{ \varphi \weak \psi \}. \\
\end{arrayeq}
Intuitively, the closure contains all possible temporal-operator sub-formulas and literals that can appear during expansion.
The locations $L$ then represent the expanded formulas, which, intuitively,  correspond to the current obligations of the system.
Thus,  the initial location will correspond to obligation $\Phi$.
Note that $L \subseteq \LTLFragT$.  
We apply simplifications to the generated formulas to ensure that $L$ is finite.
Since by definition $\mathit{cl}(\Phi)$ is finite,  we can ensure that $|L| \leq 2^{|\mathit{cl}(\Phi)|}$.

In the construction of the  initial location and the transition function we use two helper functions, 
$\mathit{introExp} : \LTLFragT \to \LTLFragT$, which performs expansion and introduces new timers, 
and $\mathit{opt} : \LTLFragT \to L$, which performs simplifications that ensure that $L$ is finite.
We let
$L_0 := \{ \mathit{opt}(\mathit{introExp}(\Phi)) \}$ 
and 
\[ \delta(\varphi, i, o, T) := (\mathit{opt}(\mathit{introExp}(\psi)),e)~\text{where}~(e, \psi) := \mathit{squeeze}(\mathit{to}(T, \mathit{tree}(\varphi, i, o))).\]
Here,  we use the additional functions
$\mathit{tree} : \LTLFragT \times \power{\mathcal{I}} \times \power{\mathcal{O}} \to \LTLFragT$, which  performs the input and outputs choices,
$\mathit{to}: \power{\timers} \times \LTLFragT \to \LTLFragT$, which handles time-outs,  and
$\mathit{squeeze} : \LTLFragT \to \effect \times \LTLFragT$, which determines remapping and reset of timers.
Below, we describe these functions in detail.\looseness=-1

{\it Remark:} Note that for  $\blnext{b}$ we use  timers of duration $b$, while for
$\beventually{b}$ and $\bweak{b}$ we use timers of duration $b + 1$.  The reason for this  is that for the latter we consider the last step as part of the timing as this simplifies the game structure.

Before describing the functions, we illustrate them on a simple example.

\begin{example}\label{ex:tg-construction}
Let $\mathcal I = \{\req\}$,  $\mathcal O = \{\gr\}$, and consider the  \LTLFrag\ formula 
$ \Phi = (\bglobally{100}\neg\gr) \land\blnext{10}(\req \to \beventually{100}\gr).$
$\Phi$ states that the system should not give a grant during the first 100 steps, 
and, if at step 10 there is a request, then a grant should be given within the following  100 steps. We show how to construct the initial location and some of the transitions in a countdown-timer game for $\Phi$. 

\noindent
{\bf Initial state $\boldsymbol{\varphi_0} = \mathit{opt}(\mathit{introExp}(\Phi))$}

The initial state is computed from $\Phi$ by expanding the formula and introducing any necessary timers. This is done by the function $\mathit{introExp}$.
The subformula $\bglobally{100}\neg\gr$ expands to $\neg \gr \land \bglobally{t_0^{101}}\neg \gr$,  reflecting the semantics of the operator $\bglobally{100}$.  This introduces the timer 
$t_0^{101}$ with duration $101$ and index $0$. 
The subformula $\blnext{10}(\req \to \beventually{100}\gr)$ expands to $\blnext{t_0^{10}}(\req \to \beventually{100}\gr)$, which introduces the timer $t_0^{10}$ for $\blnext{10}$.
The durations $101$ and $10$ of the timers correspond to the respective bounds in $\bglobally{100}$ and $\blnext{10}$,  and the index $0$ is the smallest index of a currently unused timer of the respective duration.
No timer is introduced at this step for $\beventually{100}$ as it is guarded by a $\lnext$ operator.
Thus,  the initial state is the expanded formula $\varphi_0 = \neg\gr\land(\bglobally{t_0^{101}}\neg \gr) \land\blnext{t_0^{10}}(\req \to \beventually{100}\gr)$.

\noindent
{\bf Determining transition $\boldsymbol{\delta(\varphi_0, \emptyset, \{\gr\}, \emptyset)} = (\varphi_1,e_1)$}

We apply $\mathit{tree}(\varphi_0,\emptyset, \{\gr\})$ which computes the effect of the input $\emptyset$ and output $\{\gr\}$ on the formula in the current step, and thus substitutes $\gr$ with $\top$ in $\varphi_0$. 
This results in $\mathit{tree}(\varphi_0,\emptyset, \{\gr\}) = \bot$, meaning  that this transition leads to location $\bot$.

\noindent
{\bf Determining transition $\boldsymbol{\delta(\varphi_0, \emptyset, \emptyset,\{t_0^{10}\})} = (\varphi_2,e_2)$}

Again, we first compute $\mathit{tree}(\varphi_0,\emptyset, \emptyset) = (\bglobally{t_0^{101}}\neg \gr) \land \blnext{t_0^{10}}(\req \to \beventually{100}\gr)$, which now substitutes $\bot$ for $g$.
To the result we apply the function $\mathit{to}$ that handles time-outs, here $\{t_0^{10}\}$, which means that the timer $t_0^{10}$ times out at the current step.  As a result, the subformula 
$\blnext{t_0^{10}}(\req \to \beventually{100}\gr)$ is replaced by $\req \to \beventually{100}\gr$, meaning that the formula $\req \to \beventually{100}\gr$ becomes part of the obligation at the next step,  since the timer $t_0^{10}$ has run out.
Thus,  we obtain 
$\mathit{to}(\{t_0^{10}\}, (\bglobally{t_0^{101}}\neg \gr) \land \blnext{t_0^{10}}(\req \to \beventually{100}\gr)) = 
(\bglobally{t_0^{101}}\neg \gr) \land (\req \to \beventually{100}\gr)$. 
After that, we apply function $\mathit{squeeze}$ that takes care of timers that might have become unused upon time-out.  This is reflected in the effect $e_2$ that resets all timers that do not appear in the current formula.  Thus,  in $e_2$ the timer $t_0^{10}$ that just timed out is mapped to $\reset$,  and the timer $t_0^{101}$ that is still present is mapped to itself. 
The final step is to apply function $ \mathit{introExp}$ that performs expansion on the current formula and introduces any new timers that might be needed. 
The subformula  $\bglobally{t_0^{101}}\neg \gr$ expands to $\neg\gr \land \bglobally{t_0^{101}}\neg \gr$.
The subformula $\req \to \beventually{100}\gr$ expands to $\req \to (\gr \vee \beventually{t_1^{101}}\gr)$, which introduces the timer $t_1^{101}$ for $\beventually{100}$.  Note that since the formula already contains the timer $t_0^{101}$ of duration $101$, the newly introduced timer $t_1^{101}$ has index $1$.  The functions $\mathit{to}$ and $\mathit{squeeze}$ ensure that the order between the indices of timers of the same duration represents the order in which these timers will time out.
After computing $\mathit{introExp}((\bglobally{t_0^{101}}\neg \gr) \land (\req \to \beventually{100}\gr))$ we obtain $\varphi_2 = \neg \gr \land (\bglobally{t_0^{101}}\neg \gr) \land (\req \to (\gr \vee \beventually{t_1^{101}}\gr))$.
\end{example}

\myparabf{Construction} We construct the sets of locations,  timers, and transitions, by exploring the reachable parts of $L$ from $L_0$. 
We describe several pruning mechanisms that we use in order to maintain the set of reachable locations small.\looseness=-1

\myparait{Construction Invariants.} 
To ensure correctness and keep the game generation efficient,  we maintain the following invariants for each reachable location:
\begin{compactenum}
   \item\label{invariant-1} For every reachable location $\varphi$ we have (1.a) all literals and bounded operators not guarded by a ``next" operator appear on the Boolean top-level,  and (1.b) all bounded operators at the top-level are instantiated with a timer.
   \item\label{invariant-2} For every duration $d$,  the values of the timers are ordered by index, i.e.\ $t_0^d < t_1^d < \dots t_j^d = \dots t_d^d = d$.
         The order is strict for timers whose value is not $d$.\looseness=-1
   \item\label{invariant-3} In location $\varphi$, for any $d$ and $i > 0$,  if $t_i^d \in \mathit{Timers}(\varphi)$, then $t_{i - 1}^d \in \mathit{Timers}(\varphi)$. 
\end{compactenum}

\smallskip
Invariant~(\ref{invariant-1}) is needed for correctness, and for ensuring that all literals that are relevant in the current step are considered, and that all relevant bounded operators are tracked by timers.
Invariant~(\ref{invariant-2}) ensures that we never need more than the available $d$ timers.  This holds since the timers are strictly ordered when running,  and once we would introduce $t_{d+1}^d$, $t_0^d$ would have timed out.
Furthermore, ordering the timers reduces the possible combinations of time-outs. 
Invariant~(\ref{invariant-3}) prevents having unused timers that are between used ones according to the  above order, thus reducing the possible combinations of equivalent locations.

\myparait{{\bf Function }$\boldsymbol{\mathit{tree}}$: Selection of Inputs and Outputs.}
The function $\mathit{tree}(\varphi, i, o)$ computes the effect of the input $i$ and output $o$ on the formula in the current step.
With invariant~(\ref{invariant-1}) it suffices to consider literals on the Boolean top-level, i.e. literals that are not sub-formulas of a temporal operator.
When assigning the literals in $\varphi$ according to $i$ and $o$,  we prune and select some ``obvious choices'' which can immediately be decided, using the fact that we are generating a game.
This pruning is an important part of our approach,  as in practice it can prune a significant portion of the possible locations.
Function $\mathit{tree}$ applies recursively a set of rules.
We now describe these rules in the order in which they are applied in each recursion step.
Figure~\ref{fig:tree-rules} provides a formal description.
\begin{figure}[t]
\begin{align}
            \mathit{tree}(c \lor \psi, i, o) &:= \sema{c \in o}\\
            \mathit{tree}(u \land \psi, i, o) &:= \bot \\
            \mathit{tree}(\psi, i, o) &:= \begin{cases}
                \mathit{tree}(\psi[c/\top]_T) &~\text{if}~c \in o \\
                \bot &~\text{if}~c \not\in o
            \end{cases} && c \in \mathit{ActL}(\psi), \lnot c \not\in \mathit{ActL}(\psi)\\
            \mathit{tree}(\psi, i, o) &:= \mathit{tree}(\psi[u/\bot]_T) &&u \in \mathit{ActL}(\psi), \lnot u \not\in \mathit{ActL}(\psi) \\
            \mathit{tree}(\psi, i, o) &:= \psi[u/\sema{u \in i}]_T &&u, \lnot u \in \mathit{ActL}(\psi) \\
            \mathit{tree}(\psi, i, o) &:= \psi[c/\sema{c \in o}]_T &&c, \lnot c \in \mathit{ActL}(\psi) 
\end{align}
\vspace{-5mm}
\caption{
Let $u \in \mathcal{I}$ and $c \in \mathcal{O}$.
For simplicity of the presentation we leave out the commutative and associative cases and negative literals. 
$\mathit{ActL}(\psi)$ denotes the set of literals appearing in the Boolean top-level of $\psi$.
The formula $\psi[\mathit{ap}/v]_T$ is obtained from $\psi$ by replacing $\mathit{ap}$ by $v \in \{\top, \bot \}$ for all occurrences of $\mathit{ap}$ at the Boolean top-level, but only there.
After each replacement we simplify the formula by doing constant folding.
$\sema{x \in X}$ is $\top$ if $x \in X$ and $\bot$ if $x \not\in X$.
}\label{fig:tree-rules}
\vspace{-5mm}
\end{figure}

\begin{compactenum}
    \item   With top-level disjunct $c$ that is output literal, the system wins by making the formula $\top$.  The opposite choice for the system can be safely pruned.
    \item   With top-level conjunct $u$  that is input literal, the environment wins by making the formula $\bot$.  The opposite choice can be safely pruned.
    \item   If an output proposition appears either with only positive or with only negative polarity, 
    it suffices for the system to pick the literal with the respective polarity,  as for the other choice the generated formula is subsumed. \looseness=-1
    \item   If an input proposition appears either with only positive polarity or only negative polarity, it suffices to consider the case where the environment picks the negated literal, as this case is strictly more difficult  to realize (i.e. one formula implies the other) and every strategy for this case works also for the other.
    \item   If no ``early decision'' or ``worst case-decision'' can be made, we apply the environment choice, as the environment moves first in the game.
    \item   If no environment choices are left, we generate the branching for the system.
\end{compactenum}

\myparait{{\bf Function }$\boldsymbol{\mathit{to}}$: Handling Time-out.}
A consequence of invariant~(\ref{invariant-2}) is that only timers with index 0, i.e., of the form $t_0^d$, can time out since the timers are ordered.
In addition, timers that do not appear inside a formula should not time out (this is enforced by $\mathit{squeeze}$) as we show later.
Note that this does not apply to timers with duration $1$ as these time out immediately.
We direct impossible time-outs to $\top$ since they do not occur.
Hence, $\mathit{to}(T, \varphi) := \top$ if for some $t_i^d \in T$ we have that 
$i \neq 0$, or 
$d > 1$ and $ t_i^d \not\in \mathit{Timers}(\varphi)$.
Otherwise,  $\mathit{to}(T, \varphi)$ is defined by applying the following transformations on all subformulas of $\varphi$ and timing out timers $t \in T$:
We transform $\beventually{t} \psi \rightsquigarrow \bot$, $\blnext{t} \psi \rightsquigarrow \psi$, and $\phi \bweak{t} \psi \rightsquigarrow \top$.
After applying $\mathit{to}$ we do constant folding as parts of the formula may become irrelevant.

\myparait{{\bf Function }$\boldsymbol{\mathit{squeeze}}$:  Determining remapping and reset of timers.}
When applying the functions $\mathit{tree}$ and $\mathit{to}$ some timers might become unused.
Hence, we have to ensure that invariant (\ref{invariant-3}) holds and, as stated in the previous paragraph, reset all timers that do not appear in the formula.
We define $\mathit{squeeze}(\varphi) := (e, \psi)$ as follows:
For each duration $d$, let $t_{i_j}^d \in \mathit{Timers}(\varphi)$ with indices $i_0 < i_1 < i_2 < \dots$ be the remaining timers with sorted indices $i_j$.
Then set $ e(t_j^d) := t_{i_j}^d$ if $i_j$ exists and $e(t_j^d) := \reset$ otherwise.
$\psi$ is obtained by replacing the timers $t_{i_j}^d$ by $t_j^d$.

\myparait{{\bf Function }$\boldsymbol{\mathit{introExp}}$: Expansion and Timer Introduction.}
The function $\mathit{introExp}$ performs the formula expansion and introduces new timers if necessary. 
The expansion guarantees that invariant~(\ref{invariant-1}) holds afterwards.
When introducing new timers, invariant~(\ref{invariant-2}) and invariant (\ref{invariant-3}) have also to be maintained.
This is achieved by assigning for each bound $b$ with associated duration $d$,
the timer with the next unused index, i.e. $t_j^d \not\in \mathit{Timers}(\varphi)$ where 
$t_0^d, \dots, t_{j-1}^0 \in \mathit{Timers}(\varphi)$.
Let $I(d) := \max\{ i \mid t_i^d \in \mathit{Timers}(\varphi)\}+ 1$ be the next unused index.
In addition, as timers $t_i^d$ with $i > d$ do not exist by invariant (2), expansions generating them are redirected to $\top$.
Hence, we define $\mathit{introExp}(\varphi) := \mathit{rd}(\mathit{iE}_I(\varphi))$
where 
$\mathit{rd}(\varphi) := \top$ if for some $i > d$ we have $t_i^d \in \mathit{Timers}(\varphi)$, and
$\mathit{rd}(\varphi) = \varphi$ otherwise.
The function $\mathit{iE}_I$ performing the expansion is defined by
\begin{arrayeq}{lllll}
        \mathit{iE}_I(l) &:= l &&
         \mathit{iE}_I(\varphi~o~\psi) &:=\mathit{iE}_I(\varphi)~o~\mathit{iE}_I(\psi) \\
        \mathit{iE}_I(\beventually{n} \varphi) & := \mathit{iE}_I(\varphi) \lor \beventually{t_{I(n + 1)}^{n + 1}} \varphi &&
        \mathit{iE}_I(\beventually{t} \varphi) & := \mathit{iE}_I(\varphi) \lor \beventually{t} \varphi\\
        \mathit{iE}_I(\blnext{n} \varphi) & := \blnext{t_{I(n)}^n} \varphi &&
        \mathit{iE}_I(\blnext{t} \varphi) & := \blnext{t} \varphi\\
        \mathit{iE}_I(\varphi \bweak{n} \psi) & := \mathit{iE}_I(\psi) \lor \mathit{iE}_I(\varphi) \land && 
        \mathit{iE}_I(\varphi \bweak{t} \psi) & := \mathit{iE}_I(\psi) \lor \mathit{iE}_I(\varphi)  \\
        & \quad~~(\varphi \bweak{t_{I(n + 1)}^{n + 1}} \psi) &&
        & \quad\quad\quad~~\land(\varphi \bweak{t} \psi) \\
        \mathit{iE}_I(\varphi \weak \psi) & := \mathit{iE}_I(\psi) \lor \mathit{iE}_I(\varphi) \land (\varphi \weak \psi),
\end{arrayeq}
where $l \in \{\mathit{ap}, \lnot\mathit{ap}\}$, $o \in \{\land, \lor\}$, $n \in \NN$ and $t \in \timers$.

\myparait{{\bf Function }$\boldsymbol{\mathit{opt}}$: Formula Simplification.} 
The function $\mathit{opt}$ ensures that the constructed set of locations $L$ is finite, by simplifying 
the formulas in order to avoid introducing infinitely many logically equivalent formulas.  
Since we must maintain the invariants, the simplification does not guarantee uniqueness modulo equivalence.  Nevertheless,  it ensures finiteness of $L$ and performs optimizations.

\myparabf{Definition of $\unsafe$ and Correctness }
To complete the construction of the countdown-timer game,  we define the set of unsafe locations as 
$ \mathit{UNSAFE}_L =\{ \bot\}$.  
The proof of the correctness theorem below is given in Appendix~\ref{sec:proofs-generation}.
\begin{theorem}\label{thm:correctness-generation}
    Let $\Phi \in \LTLFrag$ and $G$ be the countdown-timer game structure constructed from $\Phi$ as described above. 
    Then there exists a system realizing $\mathcal{L}(\Phi)$  if and only if the system wins in the  countdown-timer game $(G, \mathit{UNSAFE}_L)$.
\end{theorem}

We augment the construction with several extensions to improve its efficiency and expand its scope.  For instance, we combine explicit expansion with timer-based implicit expansion, which allows us to handle directly operators like single $\lnext$.  We also use approximation to handle simple assumptions of the form 
$\globally \psi$ where $\psi$ is fully bounded, i.e.,  without $\weak$.  Details can be found in Appendix~\ref{sec:extensions}.

%% file: tg-solving.tex
We now describe the second phase of our synthesis algorithm, namely the solving of the countdown-timer game generated from the $\LTLFrag$ specification. 
In a countdown-timer game, the durations of the timers, which correspond to the bounds of the temporal operators in the specification, are encoded in binary.
Hence, the set $\timerspace$ of timer valuations and thus also the safety game defined in Section~\ref{sec:timer-games}  grow exponentially in the size of the countdown-timer game.
Since our goal is to efficiently solve countdown-timer games with large durations, explicitly constructing and solving the semantic safety game is not desired.
We note, however,  that in the worst case it is not possible to avoid this blowup.
This is stated in the next theorem,  the proof of which is given in Appendix~\ref{sec:proofs-solving}.
\begin{theorem}\label{thm:complexity}
    Solving countdown-timer games is \exptime-complete.
\end{theorem}

This means that solving countdown-timer games efficiently requires an approach that manipulates sets of timer valuations symbolically,  in order to avoid, if possible,  explicit enumeration.
We propose a symbolic algorithm for solving countdown-timer games that additionally employs an iteratively refined approximation.  The method is applicable to generic  symbolic representations of the set of timer valuations.
We present an instantiation of the method with a representation composed of intervals of timer values and partial orders on timers.

\myparabf{Symbolic Game Solving}
The standard way to solve a safety game is to compute the set of states from which the environment can enforce reaching an unsafe state, and check if it intersects with the set of initial states.  If this is the case, then the environment wins the game, and otherwise the system wins.

For a  game $(G,\unsafe)$ with  $G = (\states, \states_0, \mathcal{I}, \mathcal{O}, \rho)$,
the set of states from which the environment can enforce reaching $\mathit{UNSAFE}$  is called \emph{environment attractor} and is defined as $\mathit{AttrE}_G(\mathit{UNSAFE}) = \{s \in \states \mid \exists \pi:\text{env. \ strategy}.\forall \sigma:\text{sys.\ strategy}.\exists k \in \NN.~\outcome{s}{\pi}{\sigma}_k \in  \mathit{UNSAFE}\}$. 
%
The environment wins the safety game if and only if $\mathit{AttrE}_G(\mathit{UNSAFE}) \cap S_0 \neq \emptyset$.

We solve the countdown-timer game by computing a symbolic representation of the  attractor of the environment player to the unsafe locations.
We assume a symbolic representation $\mathit{Rep}$ of the space of timer valuations $\power{\timerspace}$. 
For each $R \in \mathit{Rep}$ we denote with  $\sema{R} \subseteq \timerspace$ the subset of $\timerspace$ represented by $R$.
We represent subsets of the state space $L \times \timerspace$ of the semantic safety game using  functions from $L \to \mathit{Rep}$ where $U \in (L \to \mathit{Rep})$ represents  $\{ (l, v) \mid v \in \sema{U(l)} \}$.

The symbolic enforceable predecessor for the environment $\mathit{CPreE}_{symb} : (L \to \mathit{Rep}) \to (L \to \mathit{Rep}) $ is defined as follows. 
For $U \in (L \to \mathit{Rep})$, we let
\[
    \mathit{CPreE}_{symb} (U) := \lambda l. \bigcup_{i \subseteq \mathcal{I}} \bigcap_{o \subseteq \mathcal{O}} \bigcup_{T \subseteq \timers} \mathit{symTrans}(\delta(l, i, o, T), T, U), \text{ where} 
\]
\[
    \mathit{symTrans}((l', e), T, U) := \mathit{inc}(\mathit{effTO}(T, \mathit{remap}(e, \mathit{effReset}(e, U(l')))))
\]
is the symbolic backward application of  transition  $\delta(l, i, o, T)$ to the target set $\sema{U(l')}$. 
The operations that $\mathit{symTrans}$ requires,   from last to first, are as follows.
\begin{compactitem}
    \item   $\mathit{inc} : \rep \to \rep$ performs the backward increment of the timers, formally,  $\sema{\mathit{inc}(R)} = \{ \lambda t.~v(t) + 1 \in \timerspace \mid  v \in \sema{R} \}$.
    \item   $\mathit{effTO} : \power{\timers} \times \rep \to \rep$ models the effect of time-outs:
    $\sema{\mathit{effTO}(T, R)} = \{ v \in \sema{R} \mid \forall t \in \timers. (t \in T \to v(t) = 0) \land (t \not\in T \to v(t) \in [1, d(t)]) \} $.
    \item   $\mathit{remap} : \effect \times \rep \to \rep$ models the effect of remapping: $\sema{\mathit{remap}(e, R)} = \{ v \in \timerspace \mid \exists v' \in \sema{R}. \forall t \in \timers ~\text{s.t.}~ e^{-1}(t)~\text{is defined}.~ v(t) = v'(e^{-1}(t)) \}$.
    \item   $\mathit{effReset} : \effect \times \rep \to \rep$ models the effect of timer resets:
    $\sema{\mathit{effReset(e, R)}} = \{ v \in \sema{R} \mid \forall t \in \timers. e(t) = \reset \to v(t) = d(t) \}$.
    Note that $e^{-1}(t)$, the timer mapped to $t$ by effect $e$ is unique, since the effect is injective for values different from $\reset$, and can thus be inverted if defined.\looseness=-1
\end{compactitem}
We also require that we can preform set operations $\cup$, $\cap$, and equality checking between elements of $\rep$, in order to perform the computation.

We  employ the symbolic enforceable predecessor operator $\mathit{CPreE}_{symb}$ to  compute a symbolic representation of the environment attractor $\mathit{AttrE}_{symb}$ as follows. 
We set $\mathit{AttrE}_{symb}^0  := (\lambda l. ~\text{if } l \in \unsafe_L \text{ then } \timerspace \text{ esle }  \emptyset)$,  and then for  $n \in \NN$ we let
$\mathit{AttrE}_{symb}^{n+1} := \mathit{AttrE}_{symb}^{n} \cup \mathit{CPreE}_{symb}(\mathit{AttrE}_{symb}^{n})$.


\begin{proposition}
If $(G_T, \mathit{UNSAFE}_L)$ is a countdown-timer game 
with  $G_T = (\timers, d, L, L_0, \mathcal{I}, \mathcal{O}, \delta)$
and the safety game 
$(G, \mathit{UNSAFE_L} \times \timerspace)$ with 
$G = (L \times \timerspace, L_0 \times \{ \lambda t.d(t) \}, \mathcal{I}, \mathcal{O}, \rho)$ 
is its semantics, then for the symbolic attractor computed above it holds
$\sema{\mathit{AttrE}_{symb}(l)} = \{v \in\timerspace \mid (l,v) \in \mathit{AttrE}_{G}\}$ for every $l \in L$.
\end{proposition}

\myparabf{Approximation of Timer Valuations}
As the symbolically represented state-space described above might still lead to exploring a large number of sets, we perform an over- and under-approximation of the attractor of explored states. 

We use a \emph{threshold $k \in \NN$} to control the precision of the abstraction.
Intuitively, when approximating for $t \in \timers$ we would like to treat exactly timer values at the ``border", i.e.  timer values in $[0, k]$ and $[d(t)- k, d(t)]$, since these matter for timeouts and resets.
Our approximations $\mathit{over}: \rep \to \rep$ and $\mathit{under}: \rep \to \rep$  treat the intermediate values $[k,d(t)-k]$ like a single value-block.  The over-approximation $\mathit{over}(R)$ adds all intermediate values if one value from $R$ is inside $[k,d(t)-k]$ and the under-approximation $\mathit{under}(R)$ removes all intermediate values if one value from $R$ is not inside.  Formally:
\begin{arrayeq}{ll}
    \mathit{approx}_k(t, I) & := (I \cap [k, d(t) - k] \neq \emptyset) \land ([k, d(t) - k] \not\subseteq I) \\
    \sema{\mathit{over}(R)} & := 
    \left\{ \lambda t.
     \begin{cases}
        v(t) \cup [k, d(t) - k] &~\text{if}~\mathit{approx}_k(t,v(t)) \\
        v(t) &~\text{otherwise}
     \end{cases}
    ~\bigg|~ v \in \sema{R} \right\} \\
    \sema{\mathit{under}(R)} & := 
    \left\{  \lambda t.
    \begin{cases}
        v(t)~\backslash~[k, d(t) - k] &~\text{if}~\mathit{approx}_k(t,v(t)) \\
        v(t) &~\text{otherwise} 
    \end{cases} ~\bigg|~ v \in \sema{R} \right\} \\
\end{arrayeq}
The  attractor computation is now done as follows:
We start with $k := 1$.
For the current $k$ we compute the environment attractor once using  under- and once using over-approximation at each symbolic state in the computation. 
If the environment wins in the under-approximation, it wins the concrete game.
If the system wins in the over-approximation, it wins the concrete game. 
If neither holds, we set $k := 2 \cdot k$ and repeat.
This always terminates since for $k > d(t)/2$ the approximations become exact, and hence, one player wins for sure.  


\begin{example}
\begin{figure}[b!]
    \begin{subfigure}[b]{0.32\textwidth}
         \centering
\begin{tikzpicture}[->,shorten >=1pt,auto,node distance=2.5cm]
  \tikzstyle{every state}=[fill=none,draw=black,text=black,inner sep=1.5pt, minimum size=16pt,thick,scale=0.7]
    \node[state] (l0) at (0,0) {$l_0$};
    \node[state] (l1) at (1,0) {$l_1$};
    \node[state] (l2) at (2,0) {$l_2$};
    \node[draw=none] (lr) at (0,-1) {$\ldots$};
   \node[draw=none] (lb) at (2,-1) {$\bot$};
   \path (l0) edge node[above] {$o$} (l1)
             (l1) edge[bend left] (l2)
   			  (l2) edge  (l1)
   			  (l2) edge node[right] {$\{t_0^{1000}\}$} (lb)
   			  (l0) edge node[right] {$\lnot o$} (lr);
\end{tikzpicture}
\vspace{-.3cm}
\caption{Countdown-timer game,  $\mathit{UNSAFE}_L = \{  \bot \}$.}
\label{fig:approx-game}
     \end{subfigure}
     \hfill
     \begin{subfigure}[b]{0.67\textwidth}
         \centering
		\begin{tabular}{l|l|l|l|l||l|l|l|l} 
 & 0 &  1  & 2   & 3 &
     4 &  \ldots  & 7  \\
 \hline\hline
$l_1$ & 
$\emptyset$ &  $\emptyset$  &  $\{1\}$   & $\{1\}$ &
$\{1\},[3,997]$ &  \ldots   & $\{1\},[3,997],\{999\}$  \\
\hline
$l_2$ & 
$\emptyset$ &  $\{0\}$  &  $\{0\}$   & $\{0\},\{2\}$ &
$\{0\},\{2\}$ &  \ldots & $\{0\},[2,998],\{1000\}$  \\
\hline
\end{tabular}
         \caption{Sets during approximate attractor computation.}
         \label{table:approx-sets}
     \end{subfigure}
     \vspace{-.6cm}
\caption{Example demonstrating the effect of approximation of timer valuations.}\label{fig:approximation}
\end{figure}
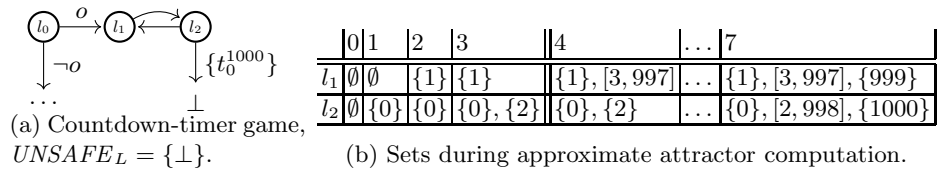
Consider a countdown-timer game,  some transitions of which are depicted in~\cref{fig:approx-game}.
From the depicted transitions, only the transition from $l_2$ to $\bot$ has a non-empty time-out set, $\{t_0^{1000}\}$.  Since the timer $t_0^{1000}$ has duration 1000, computing $\mathit{AttrE}_{symb}$  for the locations $l_1$ and $l_2$ precisely would require 1000 iterations. 
Employing over-approximation with threshold $k=3$, on the other hand,  reaches a fixed point in 7 iterations, as shown in~\cref{table:approx-sets}.  
This is helpful in cases like the one in the game in~\cref{fig:approx-game}, where the choice of transition in location $l_0$ is controlled by the system (via the output $o$). Here, the overapproximation allows the solving algorithm to quickly determine that the choice of transition to $l_1$ is loosing, while the system can win via the alternative transition.
\end{example}

\myparabf{Symbolic Representation using Boxes}
As a symbolic domain we chose an interval representation augmented with  partial orders over timers $\mathit{Rep} := \power{\mathit{PartialOrder}(\timers) \times \power{\mathit{Rec}}}$ where $\mathit{Rec} := \{\ i \in (\timers \to \NN \times \NN) \mid \forall t \in \timers, (a,b) = i(t). 0 \leq a \leq b \leq d(t) \}$ are the intervals in the form of a hyper-cube.
Intuitively, we have a set of partial-orders and for each of them we have a set of hyper-cubes.
Formally:
    \[ \sema{R} := \bigcup_{(p, C) \in R} \left(\{v \in \timerspace \mid \forall (t_1 \sim t_2) \in p:v(t_1) \sim v(t_2)\}
    \cap \bigcup_{r \in C} \lambda t. [r(t)_1, r(t)_2]\right) \]
where $r(t)_i$ is the $i$-th projection of $r(t)$.
It remains to define the necessary operations:
$\mathit{inc}$, $\mathit{effReset}$, $\mathit{effTO}$, and $\mathit{remap}$ are mostly straightforward according to their definition, as they can be performed by modifying and inspecting all intervals individually or just reordering timers.
Additionally, $\mathit{effReset}$ uses the partial order to derive bounds on timers that are in relation with a timer that is reset.
$\mathit{effTO}$ refines the partial order, since on time-out $T$, all timers in $T$ are smaller than $\timers \backslash T$.
Also the approximations can be performed point-wise on the intervals, as an approximate interval is again an interval.
As equality we use syntactic equality, i.e.\ we have to avoid redundant states to avert state space explosion.
$\cap$ is computed by using De Morgan's law and then computing the union of intervals and partial orders. 
$\cup$ can be computed by standard set union. 
We merge same partial orders and adjacent hyper-cubes to avoid redundant states.
Although this cannot remove all redundancy, it works well enough in practice.
\looseness=-1

We chose this domain since it is simple,  and, at the same time, due to the use of partial orders, well suited for the type of problem we are solving.
Our solving algorithm is generic and can accommodate other,  more sophisticated domains.
\looseness=-1

%% file: evaluation.tex
\input{evaluation-tables.tex}

We implemented\footnote{\scriptsize Available at: \url{https://github.com/phheim/lisynt}} and evaluated our approach.
We compare our prototype implementation  to 
\texttt{ebr-ltl-synth} 
 introduced in \cite{CimattiGGMT20} which performs synthesis for $\LTLEBR$.
We also compare to the state-of-the-art LTL synthesis tool \texttt{strix} version 21.0.0
\cite{DBLP:journals/acta/LuttenbergerMS20 , DBLP:conf/cav/MeyerSL18}.
In the following, we present the benchmarks we used,  the experiments, and the results.
We ran all experiments on an Intel Core i7-1165G7 processor with 16GB RAM and  a single core available. 
All times are wall-clock times.
A detailed description of the benchmarks is given in Appendix~\ref{sec:benchmarks}. 

\plotsAccumulated

\myparait{Bounded Response Benchmarks}
In our first set of  experiments we evaluate the tools on $\LTLEBR$ formulas from \cite{CimattiGGMT20},  and on 23 SYNTCOMP 2021 benchmarks\footnote{\scriptsize \url{https://github.com/SYNTCOMP/benchmarks}} that fall into $\LTLEBR$ and are used for a similar comparison in~\cite{CimattiGGMT20}.
Figure~\ref{fig:ebr-benchs} and Figure~\ref{fig:syntcomp-benchs} show the runtimes with a time-out of one minute, respectively.
Unfortunately, for roughly half of the benchmarks from~\cite{CimattiGGMT20} \texttt{strix} did not accept the input formula for being too long, since the bounded operators must be expanded explicitly upon input.
We therefore left  \texttt{strix} out for this comparison.
Figure~\ref{fig:ebr-benchs} shows that on the benchmarks from~\cite{CimattiGGMT20} both our implementation and \texttt{ebr-ltl-synth} have roughly the same runtime, ignoring different startup times. 
Figure~\ref{fig:syntcomp-benchs} shows that on the selected SYNTCOMP benchmarks all three tools are comparable.\looseness=-1

These experiments evaluate our implementation on relevant benchmarks that are partially not designed in the spirit of the problem that our approach targets. The results show that our implementation is comparable to existing tools.

\tableData

\myparait{Adaption of Real-Time Benchmarks}
In our second set of experiments, we took MTL synthesis problems from~\cite{HofmannS21} and adapted them to  $\LTLFrag$ formulas.
The benchmarks include a conveyor belt (conv-belt), a robot camera (robo-cam), and several parametrized instances of a multiple railroad-crossings controller (rail).
We discretized the real-time bounds. 
The benchmarks use up to 19 propositions and 16 bounded operators, and bounds between 60 and 4000.
Detailed results can be found in Table~\ref{tab:data}.
\texttt{ebr-ltl-synth} was not applicable to these benchmarks as we had to use assumptions (which cannot be captured by the specifications in the \LTLEBR\ fragment) to model the timed environment.\looseness=-1

These experiments show that \LTLFrag\ can  express interesting requirements  from the real-time domain by appropriate discretization.
We did not compare directly to the tool in~\cite{HofmannS21}, as the underlying modeling formalism is different, and hence we adapted the benchmarks.
However, a superficial comparison of our results to those in~\cite{HofmannS21} shows that our tool compares well (and is in some cases better). 
Furthermore, on these benchmarks our tool clearly outperforms \texttt{strix}.
\looseness=-1

\myparait{Office Robot Benchmarks}
Our last set of experiments considers benchmarks we created ourselves.
They consists of a number of specifications describing tasks for a robot in an office building with four rooms.
The benchmarks are parametrized by the number of rooms that have to be serviced.
They use up to 11 propositions and 14 bounded temporal operators, and bounds between 10 and 21600.
\looseness=-1
Detailed results can be found in Table~\ref{tab:data}.
\texttt{ebr-ltl-synth} was either not applicable due to use of assumptions (4 benchmarks) or timed out (25 benchmarks).

The results show that $\LTLFrag$ can express meaningful synthesis tasks, and that our approach is viable for solving them.
Furthermore, they show that our method indeed fulfills its purpose: for specifications requiring large bounds in the temporal operators our method clearly outperforms the state-of-the-art tools.
\looseness=-1

\myparait{Overall Analysis}
Table~\ref{tab:data}  shows that the countdown-timer game generation is very efficient compared to the solving.
As we expect to be able to improve the solving by more sophisticated symbolic techniques,  we expect the countdown-timer game based approach to be viable for even more complex properties.
In most cases the solver terminated with a low approximation threshold,  which shows the usefulness of approximation.
In our experience,  without approximation solving the benchmarks with large bounds becomes infeasible with our current technique.
\looseness=-1

%% file: evaluation-tables.tex
\newcommand{\plotsAccumulated}{%
\begin{figure}[t!]
\centering
\begin{minipage}[t]{0.41\textwidth}
\centering
\begin{tikzpicture}
\begin{axis}[ymode=log, xlabel=Accumulated Instances, ylabel=Time(ms), width=\textwidth, height=3.4cm, ylabel near ticks]
\addplot[., red] table[x=n,y=time,col sep=comma] {data/brbenchs-lisynt-times.csv}; 
\addplot[., blue] table[x=n,y=time,col sep=comma] {data/brbenchs-ebr-ltl-synth-times.csv}; 
\end{axis}
\end{tikzpicture}
\captionof{figure}{Execution times in milliseconds on the benchmarks~\cite{CimattiGGMT20}.}\label{fig:ebr-benchs}
\end{minipage}
\hfill
\begin{minipage}[t]{0.55\textwidth}
\centering
\begin{tikzpicture}
\begin{axis}[ymode=log, xlabel=Accumulated Instances, width=0.7\textwidth, height=3.4cm, ylabel near ticks, legend pos=outer north east]
\addplot[., red] table[x=n,y=time,col sep=comma] {data/scb-lisynt-times.csv}; 
\addplot[., blue] table[x=n,y=time,col sep=comma] {data/scb-ebr-times.csv}; 
\addplot[., brown] table[x=n,y=time,col sep=comma] {data/scb-strix-times.csv}; 
\addlegendentry{our tool}
\addlegendentry{ebr}
\addlegendentry{strix}
\end{axis}
\end{tikzpicture}
\captionof{figure}{Execution times in milliseconds on the $\LTLEBR$ SYNTCOMP benchmarks.}\label{fig:syntcomp-benchs}
\end{minipage}
\vspace{-6mm}
\end{figure}}

\newcommand{\winner}[1]{\textbf{#1}}
\newcommand{\tableData}{%
\begin{table}[t!]
\centering
\scriptsize
\begin{minipage}[t]{0.49\textwidth}
\centering
\begin{tabular}[t]{l|rrr|rc|r||r}
Name&$|L|$&$|\timers|$&$\tau_\mathit{Gen}$&$k$&Win.&$\tau_\Sigma$&$\tau_{\texttt{strix}}$\\\hline
$\mathit{Clean}(1)$&8&2&0.01&1&S&\winner{0.01}&3.56\\
$\mathit{Clean}(2)$&16&3&0.02&1&S&\winner{0.03}&7.99\\
$\mathit{Clean}(3)$&41&4&0.06&8&S&\winner{0.33}&21.4\\
$\mathit{Clean}(4)$&123&5&0.22&8&S&\winner{1.45}&97.3\\\hline
$\mathit{Clean}_C(1)$&10&4&0.03&1&S&\winner{0.05}&189\\
$\mathit{Clean}_C(2)$&22&5&0.08&16&S&\winner{617}&TO\\
$\mathit{Clean}_C(3)$&61&6&0.32&-&-&TO&TO\\
$\mathit{Clean}_C(4)$&205&7&1.30&-&-&TO&TO\\\hline
$\mathit{Coffee}(1)$&14&4&0.03&1&S&\winner{0.04}&TO\\
$\mathit{Coffee}(2)$&44&5&0.12&2&S&\winner{0.33}&TO\\
$\mathit{Coffee}(3)$&175&6&0.55&2&S&\winner{3.53}&TO\\
$\mathit{Coffee}(4)$&418&7&1.34&2&S&\winner{10.2}&TO\\\hline
conv-belt&9&3&0.01&1&S&\winner{0.02}&F\\
robo-cam&22&5&0.04&1&S&\winner{0.19}&F\\
rail(2,2)&647&6&2.60&1&S&\winner{3.93}&TO\\
rail(2,4)&647&6&2.58&1&S&\winner{4.05}&TO\\
rail(2,8)&647&6&2.62&1&S&\winner{3.97}&TO\\
rail(4,4)&647&7&2.67&1&S&\winner{4.10}&TO
\end{tabular}
\end{minipage}%
\hfil
\begin{minipage}[t]{0.49\textwidth}
\centering
\begin{tabular}[t]{l|rrr|rc|r||r}
Name&$|L|$&$|\timers|$&$\tau_\mathit{Gen}$&$k$&Win.&$\tau_\Sigma$&$\tau_\mathit{\texttt{strix}}$\\\hline
$\mathit{Clean}_H(1)$&3&2&0.02&512&E&\winner{0.07}&1.61\\
$\mathit{Clean}_H(2)$&3&2&0.02&512&E&\winner{0.07}&2.63\\
$\mathit{Clean}_H(3)$&3&2&0.02&512&E&\winner{0.07}&4.99\\
$\mathit{Clean}_H(4)$&3&2&0.02&512&E&\winner{0.07}&5.64\\\hline
$\mathit{Clean}_N(1)$&23&4&0.07&1&S&\winner{0.12}&TO\\
$\mathit{Clean}_N(2)$&32&4&0.10&1&S&\winner{0.27}&TO\\
$\mathit{Clean}_N(3)$&48&4&0.15&8&S&\winner{7.47}&TO\\
$\mathit{Clean}_N(4)$&75&4&0.26&8&S&\winner{13.7}&TO\\\hline
$\mathit{Coffee}_C(1)$&46&6&0.16&1&S&\winner{0.88}&F\\
$\mathit{Coffee}_C(2)$&151&7&0.59&1&S&\winner{5.51}&F\\
$\mathit{Coffee}_C(3)$&613&8&2.73&1&S&\winner{62.9}&F\\
$\mathit{Coffee}_C(4)$&1634&9&6.82&1&S&\winner{191}&F\\\hline
rail(4,8)&647&7&2.53&1&S&\winner{3.96}&TO\\
rail(8,8)&647&7&2.60&1&S&\winner{4.03}&TO\\
rail(1,1,1)&3111&7&27.8&-&-&TO&TO\\
rail(2,1,1)&9179&9&89.1&1&S&\winner{220}&TO\\
rail(2,2,2)&9179&9&93.7&1&S&\winner{225}&TO\\
\end{tabular}
\end{minipage}
\vspace{2mm}
\caption{
Results on the office-robot and adapted real-time benchmarks.
$|L|$ and $|\timers|$ are the numbers of locations and timers in the generated countdown-timer game.
$\tau_\mathit{Gen}$ is the runtime of the game generation in seconds.
$k$ is the approximation threshold on which the solving terminated.
Win. shows whether the system (S) or the environment (E) wins.
$\tau_\Sigma$ is the total runtime including the game generation and solving, where TO means a time-out after 15 minutes.
$\tau_{\texttt{strix}}$ is the runtime of \texttt{strix}.
For some benchmarks \texttt{strix} rejects the input for being too long (F) which is due to expanding the bounded operators when using \texttt{strix}. 
}\label{tab:data}
\vspace{-7mm}
\end{table}}

%% file: conclusion.tex
We presented a new synthesis approach for specifications expressed in an extension of \safetyLTL\ with bounded temporal operators. 
A distinguishing feature of our method is that it is specifically targeted at efficiently solving the synthesis problem for specifications with bounded temporal operators, in particular  those with large bounds.  Our evaluation results show that our technique performs very well on a range of benchmarks featuring such timing requirements.  The key to this success is a novel translation to a safety game with symbolically represented bounds,  whose efficiency is due to the use of effective pruning techniques. 
We observe that our method for solving the generated game is viable,  as shown by the evaluation. However,  it has potential for further  improvement by employing more performant symbolic representations and abstraction techniques.

%% file: appendix.tex
\section{Extended Techniques and Analysis}\label{sec:extensions}
\input{appendix/extensions}

\section{Proofs}\label{sec:proofs}

\input{appendix/proof-generation-correctnes}

\input{appendix/proof-tg-complexity}

\section{Example Countdown-Timer Game}
\input{appendix/example-tg}

\section{Benchmarks}\label{sec:benchmarks}
\input{appendix/benchmarks}

%% file: appendix/extensions.tex
We now give an overview of several techniques that we employ in our countdown-timer generation method that are important for its efficiency in practice.
\looseness=-1

\smallskip
\noindent
{\it Pruning with Partial Order on Timers.} 
We enhance the generation by tracking a partial order between timers of different duration. 
If two timers $t_a$ and $t_b$ with $d(t_a) < d(t_b)$ are introduced at the same time, we know that $t_b$ cannot time out before $t_a$. 
We use this information in function $\mathit{to}$ to prune more spurious time-outs.
This technique can reduce the size of the generated countdown-timer game significantly.
For example, without this technique,  in the formula $\globally(u \to (\blnext{100} c \lor \beventually{500} \lnot c))$ with $u \in \mathcal{I}$, $c \in \mathcal{O}$ it is possible for the timer of the bound 500 to time out before the timer of the bound 100 does. 
Repeated, this would lead to all timers for bound 100 being used, although only a single one is necessary.
\looseness=-1

\smallskip
\noindent
{\it Small note on Complexity.}
In the worst case, for some formula $\varphi$ we might have to generate timers for all possible indices,   which are exponentially many in the size of the formula $|\varphi|$.
Hence, $|L|$ might be triply exponential in $|\varphi|$, and $|\timerspace|$ can be doubly exponential in $|\varphi|$.
The state space $L \times \timerspace$ of the semantic safety game is triple exponential in $|\varphi|$, and hence also the worst-case solving time.
For formulas $\varphi$ for which the generated timers only have maximum indices that are logarithmically bounded w.r.t.\ their duration, $|\timerspace|$ is only exponential in $|\varphi|$ at worst.
\looseness=-1

\smallskip
\noindent
{\it Hybrid Expansion.}
Generating all timers for a bound can have  negative impact on the performance of solving  the resulting game,  due to large number of timers.
A remedy for this is to combine explicit expansion and timer-based implicit expansion.
To this end, we allow top-level temporal operators with numerical bounds,  not just timers.
Then $\mathit{introExp}$ has to be modified. For example\looseness=-1

$\mathit{iE'}_I(\blnext{n} \varphi) :=
\begin{cases}
    \blnext{t_{I(n)}^n} \varphi & \lnot\mathit{Expand explicit} \\
    \blnext{n - 1} \varphi & n > 1, \mathit{Expand explicit} \\
    \mathit{iE'}_I(\varphi) & n = 1, \mathit{Expand explicit}
\end{cases}
$

Furthermore, this allows for handling directly operators like single $\lnext$.
In light of the complexity analysis above, we expand explicitly those bounds which generate more timers than their logarithm.
This does not change the complexity class, but reduces $|\timerspace|$ exponentially,  with limited impact on the size of $L$, which improves the overall performance in practice.

\smallskip
\noindent
{\it Assumptions in \LTLFrag.}
Since \LTLFrag\ does not include unbounded ``eventually" operators,  we cannot express \emph{assumptions of the form $\globally \psi$}.  
Due to this,  we use an approximation to be able to handle simple assumptions of the form 
$\globally \psi$ where $\psi$ is fully bounded, i.e.,  without $\weak$.
The idea is to expand the assumptions separately along with the other formulas. 
If they become equivalent to $\bot$, we know for sure a violation has occurred and the system wins.
Otherwise they do not influence the outcome of the game.
This approximation is sound, i.e. if the system wins with approximation it would also win without, but not complete, i.e. if the system loses in the approximation we cannot make any statement.

%

%% file: appendix/proof-generation-correctnes.tex
\subsection{Proofs from \Cref{sec:construction}}\label{sec:proofs-generation}

\newcommand{\assoc}[2]{\langle#1, #2\rangle}
\newcommand{\explaw}{\mathit{expand}}
\newcommand{\constfold}{\mathit{cfold}}
\newcommand{\ap}{\mathit{ap}}
\newcommand{\traces}{\ensuremath{{\left(\power{\mathcal{I} \cup \mathcal{O}}\right)}^\omega}}
\newcommand{\Iff}{~\Leftrightarrow~}

\setcounter{theorem}{0}

\begin{theorem}
    Let $\Phi \in \LTLFrag$ and $G$ be the countdown-timer game structure constructed from $\Phi$ as described in~\Cref{sec:construction}.
    Then there exists a system realizing $\mathcal{L}(\Phi)$  if and only if the system wins in the  countdown-timer game $(G, \mathit{UNSAFE}_L)$.
\end{theorem}
\begin{proof}
The idea of the proof is to identify the semantic safety game of the given countdown game with expansions of $\Phi$ as $\LTLFrag$ formulas.
Using these expansion we can then show that the safety game coincides with $\mathcal{L}(\Phi)$.

Let $G = (\timers, d, L, L_0, \mathcal{I}, \mathcal{O}, \delta)$ and let $(S, S_0,  \mathcal{I}, \mathcal{O}, \rho)$ be the game structure of the semantic safety game of $G$ according to~\Cref{def:timer-game-semantics}.

Note that $S$ has the form $S = \mathit{PositiveBooleanCombinations}(\mathit{cl} (\Phi)) \times \timerspace$.
We associate each state in $S$ with a formula in $\LTLFrag$ that is yielded by inserting the timer values as respective bounds.
Intuitively, we reverse the abstraction of the timers. 
We define $\assoc{\cdot}{\cdot} : \LTLFragT \times \timerspace \to \LTLFrag$ as
\begin{align*}
    \assoc{l}{v} &:= l &~\text{for}~l \in \{\ap, \lnot \ap\} \\
    \assoc{\varphi~o~\psi}{v} &:= \assoc{\varphi}{v}~o~\assoc{\psi}{v} &~\text{for}~o \in \{\land, \lor\} \\
    \assoc{\blnext{t} \varphi}{v} &:= \blnext{v(t)} \assoc{\varphi}{v} \\
    \assoc{\beventually{t} \varphi}{v} &:= \beventually{v(t)- 1} \assoc{\varphi}{v} &~\text{if}~v(t) > 0\\
    \assoc{\beventually{t} \varphi}{v} &:= \bot &~\text{if}~v(t) = 0\\
    \assoc{\varphi \bweak{t} \psi}{v} &:= \assoc{\varphi}{v} \bweak{v(t) - 1} \assoc{\psi}{v} &~\text{if}~v(t) > 0 \\
    \assoc{\varphi \bweak{t} \psi}{v} &:= \top &~\text{if}~v(t) = 0 \\
    \assoc{\varphi \weak \psi}{v} &:= \assoc{\varphi}{v} \weak \assoc{\psi}{v}
\end{align*}

Furthermore, our argument is based on explicit expansion of $\LTLFrag$ formulas. Therefore, we define as single expansion and insertion set as step $\explaw: \LTLFrag \times (\mathcal{I} \cup \mathcal{O}) \to \LTLFrag$ as
\begin{align*}
    \explaw(\ap, m) &:= \begin{cases} \top &~\text{if}~\ap \in m \\ \bot &~\text{if}~\ap \not\in m \end{cases} \\
    \explaw(\lnot\ap, m) &:= \begin{cases} \bot &~\text{if}~\ap \in m \\ \top &~\text{if}~\ap \not\in m \end{cases} \\
    \explaw(\varphi~o~\psi, m) &:= \constfold(\explaw(\varphi, m)~o~\explaw(\psi, m)) & o \in \{\land, \lor \} \\
    \explaw(\blnext{0} \varphi, m) &:= \explaw(\varphi,m) \\
    \explaw(\blnext{n} \varphi, m) &:= \blnext{n-1} \varphi &~\text{if}~n > 0\\
    \explaw(\beventually{0} \varphi, m) &:= \explaw(\varphi,m)\\
    \explaw(\beventually{n} \varphi, m) &:= \constfold(\explaw(\varphi, m) \lor \beventually{n-1} \varphi) &~\text{if}~n > 0 \\
    \explaw(\varphi \bweak{0} \psi, m) &:= \constfold (\explaw(\psi, m) \lor \explaw(\varphi, m)) \\
    \explaw(\varphi \bweak{n} \psi, m) &:= \constfold(\explaw(\psi, m) \lor \explaw(\varphi, m) \land (\varphi \bweak{n - 1} \psi)) &~\text{if}~n > 0 \\
    \explaw(\varphi \weak \psi, m) &:= \constfold(\explaw(\psi, m) \lor \explaw(\varphi, m) \land (\varphi \weak \psi)) 
\end{align*}
where constant folding is defined as
\begin{align*}
    \constfold(\top \land \psi) = \constfold (\psi \land \top) = \psi \\
    \constfold(\bot \land \psi) = \constfold (\psi \land \bot) = \bot \\
    \constfold(\bot \lor \psi) = \constfold (\psi \lor \bot) = \psi \\
    \constfold(\top \lor \psi) = \constfold (\psi \lor \top) = \top
\end{align*}
Furthermore, for a trace $\tau \in \traces$, expansion up to step $k$ is defined as inductively as
\begin{align*}
    \explaw^0(\varphi, \tau) & := \varphi \\
    \explaw^k(\varphi, \tau) & := \explaw(\explaw^{k-1}(\varphi, \tau), \tau_{k-1}) &~\text{for}~k > 0
\end{align*}

For expansion, we can show the following (the proof follows later).
\begin{lemma}\label{lem:equivalence-expansion-unsat}
    For any trace $\tau \in \traces$, and formula $\varphi \in \LTLFrag$:
        \[ \tau \not\vDash_0 \varphi  \Leftrightarrow \exists k \in \NN. \explaw^k(\varphi, \tau) = \bot \]
\end{lemma}

We now link expansion to the generated games. 
We therefore assume for now that $\mathit{tree}$ does no pruning, i.e. it only replaces Boolean top-level literals according to their values and does constant folding.
We argue later why the pruning is sound. 

Let $\equiv$ define equivalence of formulas on the Boolean top-level, then
\begin{lemma}\label{lem:horror}
    For any $v \in \timerspace$, reachable $\varphi \in L$, $u \in \power{\mathcal{I}}$, and 
    $c \in \power{\mathcal{O}}$
        \[\langle\rho((\varphi, v), u, c)\rangle \equiv \explaw(\assoc{\varphi}{v}, u \cup c)\]
\end{lemma}
The proof is given later.

We define the $k$-iteration of $\rho$ for $k \in \NN$ on $s_0 \in S$, $i \in {\left(\power{\mathcal{I}}\right)}^\omega$, and $o \in {\left(\power{\mathcal{O}}\right)}^\omega$ as follows:
\begin{align*}
    \rho^0(s_0, \tau) & := s_0 \\
    \rho^k(s_0, \tau) & := \rho(\rho^{k-1}(s_0, \tau), i_{k-1}, o_{k-1}) &~\text{for}~k > 0
\end{align*}

For any $v \in \timerspace$, reachable $\varphi \in L$, $i \in {\left(\power{\mathcal{I}}\right)}^\omega$, and $o \in {\left(\power{\mathcal{O}}\right)}^\omega$, iterating~\Cref{lem:horror} yields
    \[ \langle\rho^k((\varphi, v), i ,o)\rangle \equiv \explaw^k(\assoc{\varphi}{v}, i \cup o) \]
for $k \in \NN$ where $i \cup o$ is defined pointwise. 
Note that this is only the case, as $\rho$ and $\explaw$ are semantic preserving, i.e. applying them to equivalent formulas provides an equivalent formula.
Combining this with~\Cref{lem:equivalence-expansion-unsat} yields 
\begin{align*}
    i \cup o \not\in \mathcal{L}(\assoc{\varphi}{v})  & \Leftrightarrow 
    i \cup o \not\vDash_0 \assoc{\varphi}{v} \\ & \Leftrightarrow
    \exists k \in \NN. \langle\rho^k((\varphi, v), i ,o)\rangle = \bot  \\ & \Leftrightarrow
    \exists k \in \NN. \rho^k((\varphi, v), i ,o) \in \{\bot \} \times \timerspace \\ & \Leftrightarrow
    \exists k \in \NN. \rho^k((\varphi, v), i ,o) \in \unsafe
\end{align*}
Note that $L_0 = \{ \mathit{opt}\mathit{introExp}(\Phi)\}$ hence $S_0$ is the singleton $S_0 = \{ s_0 \}$ with $s_0 = (\mathit{opt}\mathit{introExp}(\Phi), \lambda t.~d(t))$.
By construction of $\mathit{introExp}$ and $\mathit{opt}$, $\Phi$ is equivalent to $\langle s_0 \rangle$, i.e. $\mathcal{L}(\Phi) = \mathcal{L}(\langle s_0 \rangle s_0)$, since $\mathit{introExp}$ only applies expansion laws and insert timers that will be re-substituted by $\assoc{\cdot}{\cdot}$ accordingly.
Hence, 
    \[ i \cup o \not\in \mathcal{L}(\Phi) \Leftrightarrow \exists k \in \NN.~ \rho^k(s_0, i ,o) \in \unsafe\]
Since $i$ and $o$ are all-quantified, we just have to lift this statement in the context of systems and strategies. 
We prove our main statement by showing that the environment wins in the game if and only if there is no implementation for $\mathcal{L}(\Phi)$.
If there is no implementation for $\mathcal{L}(\Phi)$ then for all systems  there exists an input sequence $i \in {\left(\power{\mathcal{I}}\right)}^\omega$ such that for the resulting output sequence $o \in {\left(\power{\mathcal{O}}\right)}^\omega$, defined by $(c_{t+1}, o_{t}) = \gamma(c_t, i_t)$ for $t \in \NN$, $i \cup o \not\in \mathcal{L}(\Phi)$. 
Using the above equivalence, $\exists k \in \NN.~ \rho^k(s_0, i ,o) \in \unsafe$, i.e. any implementation looses the safety game.
Since safety games are solvable if and only if there are solvable with memory-less strategies the environment always wins the safety game.
Hence, the equivalence holds.

We now prove our two lemmas:
\begin{proof}[\Cref{lem:equivalence-expansion-unsat}]
Let $\tau[n] := \lambda t.~\tau_{t + n}$ be $\tau$ starting at position $n$.
We prove the more general statement
    \[ \forall t \in \NN.~\tau \not\vDash_t \varphi  \Iff \exists k \in \NN. \explaw^k(\varphi, \tau[t]) = \bot \]
by structural induction over $\varphi$:
\begin{description}
\item[$\ap$:] \[\ap \not\vDash_t \tau \Iff \ap \not\in \tau_t \Iff \explaw(\ap, \tau[t]) = \bot \Iff \explaw^1(\ap, \tau[t]) = \bot\]
\item[$\lnot\ap$:] \[\lnot\ap \not\vDash_t \tau \Iff \ap \in \tau_t \Iff \explaw(\lnot\ap, \tau[t]) = \bot \Iff \explaw^1(\lnot\ap, \tau[t]) = \bot\]
\item[$\varphi \land \psi$:]
\begin{arrayeq}{rll}
    & \tau \not\vDash_t \varphi \land \psi \\
\Iff& (\tau \not\vDash_t \varphi) \lor (\tau \not\vDash_t \psi) \\
\Iff& (\exists k \in \NN.~\explaw^k(\varphi, \tau[t]) = \bot) \\ & \lor  (\exists k \in \NN.~\explaw^k(\psi, \tau[t]) = \bot) &~\text{by IH} \\
\Iff& \exists k \in \NN.~\explaw^k(\varphi \land \psi, \tau[t]) = \bot &~\text{due to } \constfold
\end{arrayeq}
\item[$\varphi \lor \psi$:]
\begin{arrayeq}{rll}
    & \tau \not\vDash_t \varphi \lor \psi \\
\Iff& (\tau \not\vDash_t \varphi) \land (\tau \not\vDash_t \psi) \\
\Iff& (\exists k \in \NN.~\explaw^k(\varphi, \tau[t]) = \bot) \\ & \land  (\exists k \in \NN.~\explaw^k(\psi, \tau[t]) = \bot) &~\text{by IH} \\
\Iff& \exists k \in \NN.~\explaw^k(\varphi \lor \psi, \tau[t]
\end{arrayeq}
Note that the last steps holds as when first disjunct becomes $\bot$ constfold removes it.
The bound before the disjunction expands to $\bot$ is the maximum out of both disjuncts.
\item[$\blnext{n} \psi$:]
\begin{arrayeq}{rll}
    & \tau \not\vDash_t \blnext{n} \psi \\
\Iff& \tau \not\vDash_{t + n} \psi \\
\Iff& \exists k \in \NN.~\explaw^k(\psi, \tau[t + n]) = \bot &~\text{by IH}\\
\Iff& \exists k \in \NN.~\explaw^k(\blnext{n} \psi, \tau[t]) = \bot &~\text{by iterating rule for}~\blnext{\cdot}~\text{in}~\explaw
\end{arrayeq}
\item[$\beventually{n} \psi$:]
\begin{arrayeq}{rll}
    & \tau \not\vDash_t \beventually{n} \psi \\
\Iff& \forall i \leq n.~\tau \not\vDash_{t + i} \psi \\
\Iff& \forall i \leq n.~\exists k \in \NN.~\explaw^k(\psi, \tau[t + i]) = \bot  &~\text{by IH} \\
\Iff&  \exists k \in \NN.~\explaw^k(\beventually{n}\psi, \tau[t]) 
\end{arrayeq}
The last step holds as $\explaw$ will expand $\beventually{n}\psi$ as disjuncts over all the $n$ steps. 
This is folded to $\bot$ if an only if all sub-parts become $\bot$.
\item[$\varphi \weak \psi$:]
\begin{arrayeq}{rll}
    & \tau \not\vDash_t \varphi \weak \psi \\
\Iff& \exists j. (\tau \not\vDash_{t+j} \varphi) \land \forall j \leq j. (\tau \not\vDash_{t+i} \psi) \\
\Iff& \exists j. (\exists k_1 \in \NN.~\explaw^{k_1}(\varphi, \tau[t + i]))  \land \forall j \leq j. (\exists k_2 \in \NN.~\explaw^{k_2}(\psi, \tau[t + i])) 
\end{arrayeq}
    Note that
\[ \explaw^k(\varphi \weak \psi, \tau) = a_0 \lor b_0 \land (a_1 \lor b_1 \land (\dots (\varphi \weak \psi)))\]
    where $a_i := \explaw^{k-i}(\psi, \tau)$ and $b_i = \explaw^{k-i} (\varphi, \tau)$ are expansion of $\varphi$ and $\psi$ over the different expansions steps.
    Now observe that this term can only be folded to $\bot$, if there is some expansion step $m$ where $b_m$ is $\bot$ and all for all $i \leq m$, $a_i$ is also $\bot$.
    Hence, the above therm is equivalent to 
        \[ \exists k \in \NN.~\explaw^k(\varphi \weak \psi, \tau[t]) = \bot\] 
    Note that $k$ is the maximum out of $k_1 + i$ and all $k_2 + j$.
\item[$\varphi \bweak{n} \psi$:] Analogous to the previous case but with a bound. If it is not satisfied then within the bound either $\varphi$ is not satisfied and $\psi$ was never satisfied. 
When the respective terms in the expansion become $\bot$ constant folding will fold everything to $\bot$.
\end{description}

~\qed
\end{proof}
\begin{proof}[\Cref{lem:horror}]
$\rho((\varphi, v), u ,c)$ is define as $(l', v')$ where
\begin{itemize}
    \item   $T := \{ t \mid v(t) - 1 \leq 0\}$
    \item   $(e, \psi) := \mathit{squeeze}(\mathit{to}(T, \mathit{tree}(\varphi, u, c)))$
    \item   $l' := \mathit{opt}(\mathit{introExp}(\psi))$.
    \item   \[v' := 
                \lambda t.~\begin{cases}
                    v(e(t)) &~\text{if}~e(t) \in \timers \\
                    d(t)    &~\text{if}~e(t) = \reset
                \end{cases}
            \]
\end{itemize}
First note that $\mathit{squeeze}$ only reorders timers such under a specific that $\assoc{\cdot}{\cdot}$ is invariant under the changes made in $v'$. 
As $\mathit{opt}$ only modifies the top-level Boolean level, it remains to show that $\mathit{introExp}$, $\mathit{tree}$ and $\mathit{to}$ perform under $\assoc{\cdot}{\cdot}$ as $\explaw$. 
First note that the time out behavior of $\mathit{to}$ corresponds to the three cases with bounded operator cases with $n = 0$ in $\explaw$ since in the timer game we use timers with incremented duration compared to the respective bound. 
$\mathit{to}$ then just applies the neural-element which is remove by constant folding.
$\mathit{introExp}$ expands the formulas according to the expansion laws like $\explaw$ for the other cases. 
By construction, the introduction of timers is equivalent under $\assoc{\cdot}{\cdot}$.
$\mathit{tree}$ assigns literals in the same cases $\explaw$ does, namely only for top-level Boolean literals.
~\qed
\end{proof}

It remains to show that the pruning made by $\mathit{tree}$ is sound, i.e does not change how is winning. 
For pruning rules (1) and (2) this is evident as in these cases the player can enforce enter their respective winning states.
For pruning rule (3) the system if forced to choose a successor $\psi_G$ over $\psi_B$, such that $\mathcal{L}(\psi_B) \subseteq \mathcal{L}(\psi_G)$. 
Hence, every strategy that looses for $\psi_G$ looses for $\psi_B$, i.e. this does not change whether the environment wins.
Analogously, for pruning rule (4) is the system loose for the selected choice $\psi_S$ of the rules the environment would just select this, if not then the system wins anyways as for the alternative choice of the environment $\psi_N$, $\mathcal{L}(\psi_S) \subseteq \mathcal{L}(\psi_N)$.\qed
\end{proof}

%% file: appendix/proof-tg-complexity.tex
\subsection{Proofs from~\Cref{sec:solving}}\label{sec:proofs-solving}
\setcounter{theorem}{1}

\begin{theorem}
    Solving countdown-timer games is \exptime-complete.
\end{theorem}

\begin{proof}

Inclusion in \exptime\ follows from the fact that building and solving the semantic safety game structure $G$ can be done in time exponential in the size of the countdown-timer game structure $G_T$. 
To see this,  first note that the number of states $n_s = |L| \cdot  |\timerspace|$ in the game $G$ can be bounded by
    \[ n_s = |L| \cdot |\timerspace| \leq 
       |L| \cdot {(\max{ d(t) ~:~ t \in \timers})}^{|\timers|} \leq 
       |G_T| \cdot {(2^{|G_T|})}^{|G_T|} \leq 2^{\poly(|G_T|)},\]
where $|G_T|$ is the encoding length of the timer-game with numbers encoded in  binary.
The number of transitions in $G$ can be bounded by 
    \[ n_t = n_s \cdot |\power{\mathcal{I}}| \cdot|\power{\mathcal{O}}| \leq
     2^{\poly(|G_T|)}.\]
Since the size of the  game structure $G$ is $n_s + n_t + |\power{\mathcal{I}}| + |\power{\mathcal{O}}| \leq 2^{\poly(|G_T|)}$,  and safety games can be solved in time linear in the size of the game graph, it follows that 
solving $(G_T,\unsafe)$  can be done in exponential time.

We show \exptime-hardness  by reduction from countdown games, introduced in~\cite{JurdzinskiSL08}.
Countdown games are two-player games in which one player wins when they can make a transition after exactly $c$ time units have elapsed, regardless of the behavior of the other player.
In~\cite{JurdzinskiSL08} it was shown that the problem of determining the winning player in countdown games is \exptime-complete.

Formally,   a countdown game is a pair 
$C= (S_C, T_C)$ where 
$S_C$ is a set of states and 
$T_C \subseteq S_C \times (\NN \setminus \{0\}) \times S_C$ is the transition relation. 
For $(s, \Delta, s') \in T$,  $\Delta$ is called the duration of the transition.
A configuration of $C$ is a pair $(s, c)$,  
where $s \in S_C$ and $c \in \NN$.
The game proceeds as follows. 
In a  configuration $(s, c)$,  
Player 1 chooses a number $\Delta$, such that 
$0 < \Delta \leq c$ and $(s, \Delta,  s') \in T_C$,  for some state $s' \in S_C$.
Then,  Player 2 chooses a transition $(s, \Delta, s') \in T_C$ 
of duration $\Delta$.
The successor configuration is $(s', c-\Delta)$. 
There are two types of terminal configurations $(s,c)$ in which no moves are available:
\begin{itemize}
\item If $c = 0$ then  $(s, c)$ is terminal and is winning for Player 1.
\item If for all transitions $(s, \Delta, s') \in T_C$ we have that $\Delta > c$,  then $(s, c)$ is terminal and winning for Player 2. 
\end{itemize}
The problem of solving countdown games is, 
given a countdown game  $C = (S_C, T_C)$ and a configuration $(s, c)$, 
with all the durations of transitions in $C$ and the number $c$  given in binary, to determine whether Player 1 has a strategy to enforce reaching a winning configuration from configuration $(s, c)$.
Deciding the winner in countdown games is \exptime-complete~\cite{JurdzinskiSL08}.
We will show how to encode this problem into the problem of solving a countdown-timer game.

Let $C = (S_C, T_C)$ be a countdown game and $(s, c)$ be its initial configuration.
We assume w.l.o.g.\ that all durations are greater than to equal to $2$ (this can be achieved by multiplying all durations and $c$ by $2$). 
We construct a timer game $(G_T, \mathit{UNSAFE}_L)$ with
$G_T = (\timers, d, L, L_0, \mathcal{I}, \mathcal{O}, \delta)$ 
such that the environment player wins $(G_T, \mathit{UNSAFE}_L)$ if and only if Player 1 wins in the countdown game. 
The game structure $G_T$ is defined as follows:
\begin{itemize} 
\item $\timers := \{g\} \cup \{ t_\Delta \mid \exists s,s'.~(s, \Delta + 1, s') \in T_C \}$ with duration $d(g) := c$ and $d(t_\Delta) = \Delta$.
The timer $g$ is used to model the global countdown value in a configuration. 
The other timers are used to model the durations of the transitions.
Note that their duration is reduced by one,  which is because we have to introduce auxiliary transitions.
\item  $L := S_C \cup T_C \cup \{ \mathit{SAFE}, \mathit{UNSAFE}\}$. 
\begin{itemize}
\item The locations $S_C$ are the respective states in the countdown game.
\item The locations $T_C$ are auxiliary states. 
In these auxiliary states,  the value of $g$ is reduced by the respective duration of the transition in the countdown game,  by staying as long as necessary in the corresponding  state (using some of the timers $t_\Delta$).
\item Locations $\mathit{SAFE}$ and $\mathit{UNSAFE}$ represent the terminal states of the game (and are used to avoid spurious choices in the transition relation).
\end{itemize}
\item  We let $\mathcal I := \mathit{Propositions}(C)$,  where  $\mathit{Propositions}(C)$ are propositions corresponding to the binary encoding of durations in $C$.  Thus, 
  \[\power{\mathcal{I}} \supseteq \{ \Delta \mid \exists s,s'.~(s, \Delta, s') \in T_C \}, \]
  and the elements of $\power{\mathcal{I}}$ represent the actions of Player 1 in $C$. 
  The spurious choices are handled by $\delta$ by redirecting to $\mathit{SAFE}$.
 \item We let $\mathcal O:= \mathit{Propositions}(T_C)$,  where  $\mathit{Propositions}(T_C)$   are propositions corresponding to the binary encoding of the transitions $T_C$. 
 Thus, 
  \[\power{\mathcal{O}} \supseteq T_C,\]
  and the elements of $\power{\mathcal{O}}$ represent the actions of Player 2 in $C$. 
 The spurious choices are handled by $\delta$ by redirecting to $\mathit{UNSAFE}$.
\end{itemize}
In $G_T$ we use the following effects:
\begin{arrayeq}{ll}
    \mathit{ea} &:= \lambda t.~\reset \\
    \mathit{eg} &:= \lambda t.~\begin{cases} g &~\text{if}~t = g \\ \reset &~\text{otherwise} \end{cases}\\
    e_\Delta &:= \lambda t.~\begin{cases} g &~\text{if}~t \in \{ g, t_\Delta \} \\ \reset &~\text{otherwise} \end{cases}
\end{arrayeq}

The transition relation $\delta$ is defined as follows:
\begin{arrayeq}{lll}
        \delta(s, i, o, \mathit{TO}) &:= (s, \mathit{ea}) &~\text{if}~s \in \{ \mathit{SAFE}, \mathit{UNSAFE}\}\\
        \delta(s, \Delta, t_C, \mathit{TO}) &:= 
        \begin{cases}
            (\mathit{SAFE}, \mathit{ea}) &~\text{if}~\lnot\exists s' \in S.~(s, \Delta,s') \in T_C\\
            (\mathit{UNSAFE}, \mathit{ea}) &~\text{otherwise if}~\lnot\exists s' \in S_C.~t_C = (s, \Delta, s')\\
            (t_C, \mathit{eg}) &~\text{otherwise if}~ g \not\in \mathit{TO}\\
            (\mathit{SAFE}, \mathit{ea}) &~\text{otherwise if}~ g \in \mathit{TO}\\
        \end{cases}&~\text{if}~s \in S_C \\
        \delta(t_C, i, o, \mathit{TO}) &:= \mathit{let}~(s, \Delta, s') := t_C~\mathit{in} \\
        & \begin{cases}
            (t_C, e_\Delta)  &~\text{if}~g, t_{\Delta-1} \not\in \mathit{TO} \\
            (s', \mathit{eg}) &~\text{if}~g \not\in \mathit{TO} \land t_{\Delta-1} \in \mathit{TO} \\
            (\mathit{SAFE}, \mathit{ea}) &~\text{if}~g \in \mathit{TO} \land t_{\Delta-1} \not\in \mathit{TO} \\
            (\mathit{UNSAFE}, \mathit{ea}) &~\text{if}~g, t_{\Delta-1} \in \mathit{TO}
        \end{cases} &~\text{if}~t_C \in T_C
\end{arrayeq}

The transition relation models the moves of Player 1 and Player 2 in $C$ as the choices of the environment player and the system player in $G_T$ respectively.

The reduction given above is polynomial. 
By construction,  the environment player wins $(G_T,\unsafe)$ if and only if Player 1 wins $C$.
Thus, we can conclude that  solving countdown-timer games is \exptime-hard.
 
This completes the proof of \exptime-completness.
\end{proof}

%% file: appendix/example-tg.tex
\begin{example}\label{ex:timer-game}
Figure~\ref{fig:timer-game} depicts a countdown-timer game encoding  an \LTLFrag\ formula specifying some simple requirements that a printer should satisfy. 
The only input proposition is $j$ (for ``job"), and the output propositions are $i$ (for ``idle"),  $w$  (for ``warmed-up") and $p$ (for ``printed"). The specification is 

$
\varphi =  i \wedge \globally\Big( \neg(w \land p) \land  
\big((\neg j \land i) \rightarrow \lnext i\big)\land
 \big((j \land i) \rightarrow \lnext((\bglobally{9} w)) \land (\neg i \buntil{299} p \land i)\big) \Big).$
This specification states that the printer cannot be simultaneously warming up and already be done with the printing,  and that it remains idle while there is no printing job.  If the printer is idle and there is a printing job, it will complete printing after at most $299$ time steps,  after a warming up phase of at least $9$ steps.\looseness=-1 \qed

\begin{figure}[h!]
\centering
\begin{tikzpicture}[->,shorten >=1pt,auto,node distance=2cm]
	\tikzstyle{every state}=[rectangle, fill=none,draw=black,text=black,inner sep=1.5pt, minimum size=12pt,thick]
	\tikzset{interim/.style = {circle, inner sep=1pt, draw=black, fill=black }}
 	\tikzset{initial text=\(\)}
	\small
	\node[state, initial] (s) {$\mathit{start}$};
	\node[interim,below of=s] (ss) {};
	\node[interim,right of=s,xshift=-.5cm] (sw){};
	\node[state,right of=sw, xshift=0cm] (w){$\mathit{warm}\text{-}\mathit{up}$};
	\node[interim,right of=w, xshift=2cm] (wp){};
	\node[state,right of=wp, xshift=1.2cm] (p) {$\mathit{print}$};
	\node[interim,below of=p,yshift=-.5cm] (pb) {};
	\node[interim,above of=w, yshift=-1cm] (ps) {};
	\node[state, below of=w, yshift=-.5cm] (bot) {$\bot$};
	\coordinate[above of=p, yshift=-1cm] (ps1) {};
	\coordinate[above of=s, yshift=-1cm] (ps2) {};

	\path (s)  edge node[above,sloped,rotate=180] {$\neg j \land i$} (ss);
	\path (ss)  edge[bend right] node[below,sloped] {reset-all} (s);
	\path (s)  edge node[above] {$j \land i$} (sw);
	\path (sw)  edge node[above] {reset-all} (w);
	\path (w)  edge node[above] {$w \land \neg i \land \neg p$} (wp);
	\path (wp)  edge[bend left] node[below, near end] {$\begin{array}{l}\{\emptyset\}: t_{10}\leftarrow t_{10}, 	\\t_{30}\leftarrow t_{300}\end{array}$} (w);
	\path (wp)  edge node[above] {$\begin{array}{ll}\{\{t_{10}\}\}: &t_{10}\leftarrow \reset, \\&t_{30}	\leftarrow t_{300}\end{array}$} (p);
	\draw (p)  edge[-] (ps1);
	\path (ps1)  edge node[above, near end,sloped] {$p \land i \land \neg w$} (ps);
	\path (ps)  edge[-] node[above, sloped,near start] {reset-all} (ps2);
	\path (ps2)  edge (s);
	\path (p)  edge node[below, sloped,rotate=180] {$\neg i \land \neg p$} (pb);
	\path (pb)  edge[bend left] node[left, near start ] {$\begin{array}{ll}\{\emptyset\}:& t_{10}\leftarrow \reset, \\&t_{30}\leftarrow t_{300}\end{array}$} (p);

	\path (s)  edge node[below] {$\neg i$} (bot);
	\path (w)  edge node[above,sloped,rotate=180] {$\neg w \lor i \lor p$} (bot);
	\path (wp)  edge node[below, sloped, near start] {$\{\{t_{300}\}, \{t_{300},t_{10}\}\}$} (bot);
	\path (p)  edge node[below, sloped, near start] {$(w \land p)\lor(\neg p \land i)$} (bot);
	\path (pb)  edge node[below, sloped] {$\{\{t_{10}\},\{t_{300}\},\{t_{10},t_{300}\}\}$} (bot);
\end{tikzpicture}
\caption{Countdown-timer game with $L = \{\mathit{start},\mathit{warm}\text{-}\mathit{up},\mathit{print}\}$. 
 	For convenience, we split each transition in two parts. 
   	The first is labeled with a Boolean formula $\alpha$ over $\mathcal I\cup\mathcal O$.
   	The second,  with a pair $T: e$, where $T \subseteq \power{\timers}$ is a set of sets of timers that time out in the given step, and $e \in \effect$ is an effect. 
	A pair of labels $\alpha$, $T:e$ connecting locations $l$ and $l'$ represents the fact that there are $(i,o) \in \power{\mathcal I} \times \power{\mathcal I}$ that satisfy $\alpha$ and $\tau \in T$, such that $\delta(l,i,o,\tau) = (l',e)$.
	We denote  $\text{reset-all}:= \power{\timers}:e_{\mathit{reset}}$ where $e_{\mathit{reset}}(t)  = \reset$ 	for all $t\in \timers$.
   For simplicity, we omit the irrelevant label elements of the transitions to the unsafe state $\bot$.  }\label{fig:timer-game}
\end{figure}
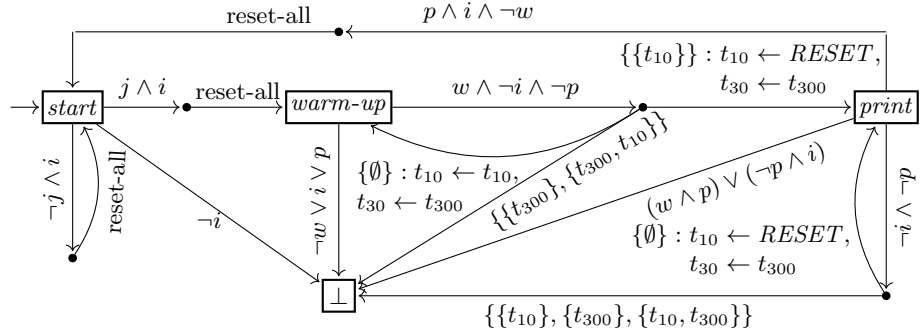
\end{example}

%% file: appendix/benchmarks.tex
\subsection{Office Robot Benchmarks}

\subsubsection{Cleaning robot for $N \in\{1,2,3,4\}$ offices}
\begin{itemize}
\item $\mathcal I = \emptyset$
\item $\mathcal O = \{\corridor,\office_1,\office_2,\office_3,\office_4\}$
\item 
$\Phi_{\mathit{Clean}(N)} := \varphi_{\mathit{start}} \wedge \varphi_{\mathit{position}} \wedge \varphi_{\mathit{mutex}}
 \wedge \varphi_{\mathit{clean-duration}} \wedge \varphi_{\mathit{clean}}^N$

\bigskip
\begin{itemize}
\item The robot starts in the corridor.
\[\varphi_{\mathit{start}}  := \corridor \]

\item The robot is in at least one location.
 \[\varphi_{\mathit{position}}  := \globally (\corridor \lor \office_1 \lor \office_2 \lor \office_3 \lor  \office_4)\]

\item The robot is in at most one location.
 \[
\begin{array}{lll} 
 \varphi_{\mathit{mutex}}  & := &
 \globally(\corridor \rightarrow (\neg\office_1 \land \neg\office_2 \land \neg\office_3 \land \neg\office_4))\\& \land&
 \globally(\office_1 \rightarrow (\neg\corridor \land \neg\office_2 \land \neg\office_3 \land \neg\office_4))\\&\land&
 \globally(\office_2 \rightarrow (\neg\office_1 \land \neg\corridor \land \neg\office_3 \land \neg\office_4))\\&\land&
 \globally(\office_3 \rightarrow (\neg\office_1 \land \neg\office_2 \land \neg\corridor \land \neg\office_4))\\&\land&
 \globally(\office_4 \rightarrow (\neg\office_1 \land \neg\office_2 \land \neg\office_3 \land \neg\corridor))
\end{array} 
\]
 
\item When staring to clean an office, clean it at least for 10 minutes.
 \[\varphi_{\mathit{clean-duration}}  := \bigwedge_{i=1}^4 \globally\big(\corridor \rightarrow \lnext (\office_i \rightarrow \bglobally{10} \office_i)\big) \]
 
 \item Start cleaning offices $1\ldots N$ every 12 hours (720 minutes).  
 \[\varphi_{\mathit{clean}}^N  := \bigwedge_{i=1}^N\globally \beventually{720} \office_i \]
\end{itemize}
\end{itemize}

\subsubsection{Cleaning and charging robot for $N \in\{1,2,3,4\}$ offices}

\begin{itemize}
\item $\mathcal I = \emptyset$

\item $\mathcal O = \{\corridor,\office_1,\office_2,\office_3,\office_4,\charge\}$

\item $\Phi_{\mathit{Clean_C}(N)} := \Phi_{\mathit{Clean}(N)} \wedge
 \varphi_{\mathit{charge-duration}} \wedge \varphi_{\mathit{charge}}$
 
\begin{itemize}

\item Formula $\Phi_{\mathit{Clean}(N)}$  is as above.

\item Charging takes 20 minutes and the charging station is in the corridor.
    \[ \begin{array}{lll} \varphi_{\mathit{charge-duration}} & := &\globally\big(\neg\charge \to \lnext(\charge \to \bglobally{20} \charge)\big)\\ &\wedge &
    \globally (\charge \to \corridor)
    \end{array}\]

\item Charge within each 6 hours (360 minutes).    
 \[\varphi_{\mathit{charge}} := \globally \beventually{360} \charge \]

\end{itemize}
\end{itemize}

\subsubsection{Cleaning robot for $N \in\{1,2,3,4\}$ offices in the presence of humans}
\begin{itemize}
\item $\mathcal I = \{\human_1,\human_2,\human_3,\human_4\}$
\item $\mathcal O = \{\corridor,\office_1,\office_2,\office_3,\office_4\}$
\item$\Phi_{\mathit{Clean_H}(N)} := \Phi_{\mathit{Clean}(N)} \wedge
 \varphi_{\mathit{no-interfere}}$
 
 \begin{itemize}

\item Formula $\Phi_{\mathit{Clean}(N)}$  is as above.

\item The robot should not be in an office if there is a human there.
 \[\varphi_{\mathit{no-interfere}}  := \bigwedge_{i=1}^4\globally( \human_i \to \neg\office_i) \]

\end{itemize}
\end{itemize}

\subsubsection{Robot cleaning during the night for $N \in\{1,2,3,4\}$ offices}
\begin{itemize}
\item $\mathcal I = \{\night\}$
\item $\mathcal O = \{\corridor,\office_1,\office_2,\office_3,\office_4\}$
\item $
\begin{array}{ll}
\Phi_{\mathit{Clean_N}(N)} := \varphi_{\mathit{night}} \to 
(&\varphi_{\mathit{start}} \wedge \varphi_{\mathit{position}} \wedge \varphi_{\mathit{mutex}} \wedge \varphi_{\mathit{clean-duration}}\wedge\\&
 \varphi_{\mathit{clean-night}} \wedge \varphi_{\mathit{clean-start-night}}^N)
\end{array}
$
 
\begin{itemize}
\item Assumption: Night comes at least every 12 hours and lasts for 12 hours.
\[\varphi_{\mathit{night}}:= \globally(\neg \night \to \beventually{720} \night) \wedge \globally(\neg \night \to \lnext(\night \to \bglobally{720} \night)) \]
\item Formulas $\varphi_{\mathit{start}}, \varphi_{\mathit{position}}, \varphi_{\mathit{mutex}},\varphi_{\mathit{clean-duration}}$ are as before.
\item  Only clean at night.
\[ \varphi_{\mathit{clean-night}}:= \bigwedge_{i=1}^4 \globally(\neg\night \to \neg\office_i)\]
\item Start cleaning on nightfall. 
\[ \varphi_{\mathit{clean-start-night}}^N:= \bigwedge_{i=1}^N \globally (\neg\night \to \lnext(\night \to \beventually{60} \office_i))\]
\end{itemize}
\end{itemize}

\subsubsection{Robot delivering coffee to $N \in\{1,2,3,4\}$ offices}
\begin{itemize}
\item $\mathcal I = \{\request_1, \request_2, \request_3, \request_4\}$
\item $\mathcal O = \{\corridor, \office_1, \office_2, \office_3, \office_4, \makecoffee\}$
\item $\Phi_{\mathit{Coffee}(N)} := \varphi_{\mathit{start}} \wedge \varphi_{\mathit{position}} \wedge \varphi_{\mathit{mutex}}
 \wedge \varphi_{\corridor}\wedge \varphi_{\mathit{coffee-making}} \wedge \varphi_{\mathit{coffee-delivery}}^N$
 \begin{itemize}
 \item Formulas $\varphi_{\mathit{start}}, \varphi_{\mathit{position}}, \varphi_{\mathit{mutex}}$ are as before.
 \item  Passing the corridor takes at least 120 seconds.
    \[ \varphi_{\corridor} := \globally(\neg\corridor \to \lnext(\corridor \to \bglobally{120} \corridor))\]
 \item The coffee machine is in office$_1$ and making coffee takes 180 seconds.
    \[ \begin{array}{lll}
    \varphi_{\mathit{coffee-making}} & := & \globally (\makecoffee \to \office_1) \\ &\land&
      \globally(\neg\makecoffee \to \lnext(\makecoffee \to \bglobally{180} \makecoffee))
    \end{array}\]
\item  Coffee requests have to be handled within 10 minutes.
     \[ \varphi_{\mathit{coffee-delivery}}^N := \bigwedge_{i=1}^N\globally (\request_{5-i} \to (\beventually{600} \makecoffee) \land (\beventually{600} \office_{5-i}))\]
 \end{itemize}
\end{itemize}

\subsubsection{Robot delivering coffee to $N \in\{1,2,3,4\}$ offices and charging}
\begin{itemize}
\item $\mathcal I = \{\request_1, \request_2, \request_3, \request_4\}$
\item $\mathcal O = \{\corridor, \office_1, \office_2, \office_3, \office_4, \makecoffee,\charge\}$
\item $\Phi_{\mathit{CoffeeC}(N)} := \Phi_{\mathit{Coffee}(N)} \wedge \varphi_{\mathit{charge-coffee}}$
 \begin{itemize}
 \item Formula $\Phi_{\mathit{Coffee}(N)}$ is as above.
     \item   Charging takes 20 minutes and has to be done after 6 hours.
    \[ \begin{array}{lll} \varphi_{\mathit{charge-coffee}} & := &
    \globally (\neg\charge \to \lnext(\charge \to (\bglobally{1200} \charge) \land (\beventually{1200} \lnext \neg\charge)))\\&\land& 
    \globally (\charge \to \lnext(\neg\charge \to (\bglobally{21600} \neg\charge) \land (\beventually{21600} \lnext \charge)))\\&\land&
    \globally (\charge \to \corridor)
    \end{array}\]     
     
 \end{itemize}
\end{itemize}

\subsection{Benchmarks Adapted From~\cite{HofmannS21}}

\subsubsection{Conveyor Belt}
This is an adaptation of Example 3 (conveyor belt) in~\cite{HofmannS21}. 
The specification is discretized in milliseconds.
\begin{itemize}
\item $\mathcal I = \{\release,  \stuck, \resume\}$
\item $\mathcal O = \{\move, \stopp\}$
\item \LTLFrag\ specification:
\[\Phi_{\text{conv-belt}} := \varphi_{\mathit{luggage}} \to  (\varphi_{\mathit{move}} \land \varphi_{\mathit{stop}} \land \varphi_{\mathit{release}} \land \varphi_{\mathit{resume}})\]
\begin{itemize}
\item Assumption: $\release$ is true if and only if luggage gets unstuck.
\[\varphi_{\mathit{luggage}}:= \globally\big((\stuck \land \lnext(\neg\stuck)) \leftrightarrow (\lnext \release)\big)\]
\item  The conveyor belt either moves or stops.
\[\varphi_{\mathit{move}}:=    \globally(\move \lor \stopp) \land \globally(\neg(\move \land \stopp))\]
\item If luggage is stuck, the conveyor belt has to stop.
\[\varphi_{\mathit{stop}}:=    \globally (\stuck \to \stopp)\]
\item After releasing, the conveyor belt has to remain stopped for 2 seconds.
\[\varphi_{\mathit{release}}:=    \globally (\release \to \bglobally{2000} \stopp)\]
\item Once the conveyor belt is released we want to resume after 3 seconds. 
(Remark: The exact behavior is unclear in the original source.) 
\[\varphi_{\mathit{resume}}:=    \globally(\resume \land \neg\stuck \to \beventually{3000} (\move \weak \stuck))\]
\end{itemize}
\end{itemize}

\subsubsection{Robot Camera}
This is an adaptation of Example 2(robot) in~\cite{HofmannS21}. 
The specification is discretized in milliseconds.
\begin{itemize}
\item $\mathcal I = \{\pick, \pput,\move\}$
\item $\mathcal O = \{\con, \coff\}$
\item \LTLFrag\ specification:
\[\Phi_{\text{robo-cam}} := \varphi_{\mathit{robot}} \to  (\varphi_{\mathit{on-off}} \land \varphi_{\mathit{switch-off}} \land \varphi_{\mathit{switch-on}} )\]
\begin{itemize}
\item Assumption: at most one of $\pick, \pput,\move$ holds and they satisfy assumption that define them.
\[\begin{array}{lll}
\varphi_{\mathit{robot}} & := & 
    \globally(\neg (\pick \land \pput)) \land
    \globally(\neg (\pick \land \move))\land
    \globally(\neg (\pput \land \move))\land \\&\land&
    \globally\big(\pick \to \lnext ((\bglobally{3000} \move) \land \beventually{3001} \pput)\big)\\& \land&
    \globally\big(\pput \to  \lnext ((\bglobally{3000} \move) \land \beventually{3001} \pick)\big)\\& \land&
    \pick
\end{array}
\]
\item  The camera is either on or off.
 \[\varphi_{\mathit{on-off}}:=   \globally (\neg (\con \leftrightarrow \coff))\]
\item The camera is not continuously on for more the 4 seconds.
 \[\varphi_{\mathit{switch-off}}:= \globally(\con \to \beventually{4000} \neg\con)\]
\item 1 second before picking or putting the camera has to be turned on.
  \[\varphi_{\mathit{swicth-on}}:=  \globally ((\beventually{1000} \pick) \to \con) \land
    \globally ((\beventually{1000} \pput)  \to \con)\]
\end{itemize}
\end{itemize}

\subsubsection{Railroad with Three Crossings $(T1,T2,T3)$}
This is an adaptation of Example 1(railroad) in~\cite{HofmannS21}.
\begin{itemize}
\item $
\begin{array}{ll}
\mathcal I = &\{\pin_1, \pin_2, \pin_3, \travel,
\closed_1, \opened_1, \transit_1, \\&
\closed_2, \opened_2, \transit_2, 
\closed_3, \opened_3, \transit_3\}
\end{array}$
\item $\mathcal O = \{\popen_1, \pclose_1, \popen_2, \pclose_2, \popen_3,\pclose_3\}$
\item \LTLFrag\ specification:
\[\Phi_{\text{rail(T1,T2,T3)}} := (\varphi_{\mathit{train}(T1,T2,T3)} \land \varphi_{\mathit{gates}}) \to  (\varphi_{\mathit{close}} \land \varphi_{\mathit{open}} )\]
\begin{itemize}
\item Assumption: Train crosses crossing 1 after $T1$ minutes,  crossing 2 after $T2$ minutes,  crossing 3 after $T3$ minutes.  After crossing, the train travels.
\[\begin{array}{lll}
\varphi_{\mathit{train}(T1,T2,T3)} & := & 
    (\bglobally{60 \cdot{T1}} \neg\pin_1 )\\ & \land &
    (\bglobally{60 \cdot{T2}} \neg\pin_2)\\ & \land &
    (\bglobally{60 \cdot{T3}} \neg\pin_3) \\ & \land &
    (\bglobally{60 \cdot{T3} +120} \neg\travel)\\ & \land &
    \opened_1 \land
    \opened_2 \land
    \opened_3 \\ & \land &
    \globally (\travel \to \lnext \travel) \\ & \land &
    \globally (\travel \to \neg(\pin_1 \lor \pin_2 \lor \pin_3)
\end{array}
    \]
    \item Assumption: behavior of the gates.
\[\begin{array}{lll}
\varphi_{\mathit{gates}} & :=  \bigwedge_{i=1}^3\Big(& 
    \phantom{\land}\globally(\neg (\opened_i \land \closed_i))\\&&\land
    \globally(\neg (\transit_i \land \closed_i))\\&&\land
    \globally(\neg (\transit_i \land \opened_i))\\&&\land
    \globally(\neg\transit_i \to \lnext(\transit_i \to \bglobally{60} \transit_i))\\&&\land
    \globally(\opened_i \land \neg\pclose_i \to \lnext \opened_i)\\&&\land
   \globally( \opened_i \land  \pclose_i \to \lnext (\transit_i \land \beventually{60} \lnext \closed_i))\\&&\land
    \globally(\closed_i \land \neg\popen_i  \to \lnext \closed_i)\\&&\land
    \globally(\closed_i \land  \popen_i  \to \lnext (transit_i \land \beventually{60} \lnext \opened_i))\Big)
\end{array}
    \]    
\item  When the train is inside, the gate is closed.
 \[\varphi_{\mathit{close}}:= \bigwedge_{i=1}^3
    \globally (\pin_i \to \closed_i)\]

\item  When the train left, the gate should open.
 \[\varphi_{\mathit{close}} :=  \bigwedge_{i=1}^3
     \globally (\travel \to \beventually{120} opened_i)
     \]
\end{itemize}
\end{itemize}

\subsubsection{Railroad with Two Crossings $(T1,T2)$}
This is an adaptation of Example 1(railroad) in~\cite{HofmannS21}.
\begin{itemize}
\item $
\mathcal I = \{\pin_1, \pin_2,  \travel,
\closed_1, \opened_1, \transit_1, 
\closed_2, \opened_2, \transit_2\}
$
\item $\mathcal O = \{\popen_1, \pclose_1, \popen_2, \pclose_2\}$
\item \LTLFrag\ specification:
\[\Phi_{\text{rail(T1,T2)}} := (\varphi_{\mathit{train}(T1,T2)} \land \varphi_{\mathit{gates}}') \to  (\varphi_{\mathit{close}}' \land \varphi_{\mathit{open}}' )\]

\begin{itemize}
\item Assumption: Train crosses crossing 1 after $T1$ minutes,  and crossing 2 after $T1+T2$ minutes. After crossing, the train travels.
\[\begin{array}{lll}
\varphi_{\mathit{train}(T1,T2)} & := & 
    (\bglobally{60 \cdot{T1}} \neg\pin_1 )\\ & \land &
    (\bglobally{60 \cdot (T1+T2)} \neg\pin_2)\\ & \land &
    (\bglobally{60 \cdot{ (T1+T2)} + 120 (\star)} \neg\travel)\\ & \land &
    \opened_1 \land
    \opened_2 \\ & \land &
    \globally (\travel \to \lnext \travel) \\ & \land &
    \globally (\travel \to \neg(\pin_1 \lor \pin_2)
\end{array}
    \]
    \item Assumption: behavior of the gates.
\[\begin{array}{lll}
\varphi_{\mathit{gates}} & :=  \bigwedge_{i=1}^2\Big(& 
    \phantom{\land}\globally(\neg (\opened_i \land \closed_i))\\&&\land
    \globally(\neg (\transit_i \land \closed_i))\\&&\land
    \globally(\neg (\transit_i \land \opened_i))\\&&\land
    \globally(\neg\transit_i \to \lnext(\transit_i \to \bglobally{60} \transit_i))\\&&\land
    \globally(\opened_i \land \neg\pclose_i \to \lnext \opened_i)\\&&\land
   \globally( \opened_i \land  \pclose_i \to \lnext (\transit_i \land \beventually{60} \lnext \closed_i))\\&&\land
    \globally(\closed_i \land \neg\popen_i  \to \lnext \closed_i)\\&&\land
    \globally(\closed_i \land  \popen_i  \to \lnext (transit_i \land \beventually{60} \lnext \opened_i))\Big)
\end{array}
    \]    
\item  When the train is inside, the gate is closed.
 \[\varphi_{\mathit{close}}:= \bigwedge_{i=1}^2
    \globally (\pin_i \to \closed_i)\]

\item  When the train left, the gate should open.
 \[\varphi_{\mathit{close}} :=  \bigwedge_{i=1}^2
     \globally (\travel \to \beventually{120} opened_i)
     \]
\end{itemize}

\end{itemize}

$\star$ For ($T1 = 2$, $T2=4$),  instead of 120 at this position we have 100,  due to a mistake we discovered too late.

\subsection{Benchmarks from~\cite{CimattiGGMT20}} 

The exact benchmarks were taken from \url{https://gitlab.fbk.eu/lgeatti/ksy-tool.git}.
They consist of four formula types $\varphi_A(N), \varphi_B(N), \varphi_C(N),\varphi_D(N)$ with $N \in \{2, \dots, 200\}$. $\varphi_A$, $\varphi_B$ are realizable and $\varphi_C$, $\varphi_D$ are unrealizable. 

$\mathcal I := \{u_0,\ldots,u_N\}$ and $\mathcal O := \{c0,\ldots, cN\}$.
\begin{align*}
    \varphi_A(N) &:= \bigwedge_{i = 0}^{N-1} (\blnext{1+2+\dots+i}\globally c_i) \land \blnext{1+2+\dots+N}\globally(c_N \lor u) \\
    \varphi_B(N) &:= \bigwedge_{i = 0}^N (\blnext{1+2+\dots+i}\globally (c_i \lor u_i) \\
    \varphi_C(N) &:= (\globally c_0) \land \bigvee_{i = 0}^N \globally \left( \bigwedge_{j=0}^i u_j \right) \\
    \varphi_D(N) &:= c_0 \land \bigwedge_{i=0}^N \blnext{i} (u_i \lor u_{i+1}) \\
\end{align*}